\documentclass[twocolumn,showpacs,preprintnumbers,amsmath,amssymb]{revtex4}

\usepackage{graphicx}
\usepackage{dcolumn}
\usepackage{bm}

\begin{document}

\preprint{APS/123-QED;}

\title{Resonant Spin-Flavor Conversion of Supernova Neutrinos:\\ 
Dependence on Electron Mole Fraction}

\author{Takashi Yoshida$^1$}
\email{tyoshida@astron.s.u-tokyo.ac.jp}
\author{Akira Takamura$^2$}%
\author{Keiichi Kimura$^3$}%
\author{Hidekazu Yokomakura$^3$}%
\author{Shio Kawagoe$^4$}%
\author{Toshitaka Kajino$^{1,5}$}%
\affiliation{
$^1$Department of Astronomy, Graduate School of Science, University of Tokyo, 
Tokyo 113-0033, Japan \\
$^2$Department of Mathematics, Toyota National College of Technology, 
Aichi 471-8525, Japan \\
$^3$Department of Physics, Graduate School of Science, Nagoya University, 
Aichi 464-8602, Japan \\
$^4$Knowledge Dissemination Unit, Institute of Industrial Science,
University of Tokyo, Tokyo 153-8505, Japan \\
$^5$National Astronomical Observatory of Japan and The Graduate University 
for Advanced Studies, Tokyo 181-8588, Japan}%

\date{\today}

\begin{abstract}
Detailed dependence of resonant spin-flavor (RSF) conversion of supernova 
neutrinos on electron mole fraction $Y_e$ is investigated.
Supernova explosion forms a hot-bubble and neutrino-driven wind region of
which electron mole fraction exceeds 0.5 in several seconds after the
core collapse.
When a higher resonance of the RSF conversion is located in the innermost 
region, flavor change of the neutrinos strongly depends on the sign of 
$1 - 2Y_e$.
At an adiabatic high RSF resonance the flavor conversion of 
$\bar{\nu}_e \leftrightarrow \nu_{\mu,\tau}$ occurs in $Y_e < 0.5$
and normal mass hierarchy or in $Y_e > 0.5$ and inverted mass hierarchy.
In other cases of $Y_e$ values and mass hierarchies, the conversion of 
$\nu_e \leftrightarrow \bar{\nu}_{\mu,\tau}$ occurs.
The final $\bar{\nu}_e$ spectrum is evaluated in the cases of
$Y_e < 0.5$ and $Y_e > 0.5$ taking account of the RSF conversion.
Based on the obtained result, time variation of the event number
ratios of low $\bar{\nu}_e$ energy to high $\bar{\nu}_e$ energy is
discussed.
In normal mass hierarchy, an enhancement of the event ratio should be
seen in the period when the electron fraction in the innermost region 
exceeds 0.5.
In inverted mass hierarchy, on the other hand, a dip of the event
ratio should be observed.
Therefore, the time variation of the event number ratio is useful 
to investigate the effect of the RSF conversion.
\end{abstract}

\maketitle

\section{\label{sec:level1}Introduction}

Core-collapse supernovae (SNe) supply a huge amount of neutrinos 
$(N_\nu \sim 10^{58})$ in a time scale of $\sim 10$ s.
If one SN explodes in our Galaxy, thousands of neutrinos are 
expected to be detected by 10 kton size neutrino detectors.
The energy spectra of SN neutrinos will provide various information
of SN explosion mechanism as well as neutrino oscillation parameters.
SN neutrinos are emitted from proto-neutron stars, where the 
density is much larger than the density of higher resonance of the
Mikheyev-Smirnov-Wolfstein (MSW) effect.
The energy spectra and the dependence on neutrino oscillation parameters
are quite different from solar neutrinos (e.g., \cite{ds00}).
Therefore, the neutrinos released from SNe in our Galaxy will supply
fruitful information on neutrino physics and astrophysics.

Although the magnetic moment of neutrinos is considered to be 
^^ ^^ very small,'' a finite magnetic moment may affect astrophysical
phenomena with ^^ ^^ very strong'' magnetic fields.
The standard model of particle physics suggested that the magnetic moment
of neutrinos is smaller than the order of $\sim 10^{-18} \mu_B$, 
where $\mu_B=e \hbar/2 m_e c$ is Bohr magnetons, $e$ is the charge of an
electron, and $m_e$ is electron mass (review in \cite{pu92}).
However, particle theories beyond the standard model have suggested
that the upper limit of the neutrino magnetic moment is up to
$\sim 10^{-10} \mu_B$ (e.g., \cite{fy87}) and grand unified theory permits 
such a ^^ ^^ large'' magnetic moment of neutrinos.
Therefore, if evidence for a neutrino magnetic moment is found, it will
become a quite new trace of particle physics beyond the standard model.
Neutrino experiments also have constrained the upper limit of
neutrino magnetic moment.
Recently, the TEXONO experiment deduced the upper limit 
$\mu_{\bar{\nu}_e} < 7.4 \times 10^{-11} \mu_B$ from $\bar{\nu}_e$ detection
\cite{wo07}.
The GEMMA experiment obtained a stronger constraint 
$\mu_{\nu_e} < 5.8 \times 10^{-11} \mu_B$ at 90 \% C.L. \cite{be07}.
Astrophysical limit has been evaluated from plasmon decays in stars
in globular clusters as $\mu_\nu < 3 \times 10^{-12} \mu_B$ \cite{ra99}.

If neutrinos are Majorana particles and have a finite magnetic moment,
they have only transition magnetic moment.
In this case, a spin precession between a left-handed neutrino $\nu_L$
and a right-handed (anti)neutrino $\bar{\nu}_R$ with different
flavors occurs in strong magnetic field (e.g., \cite{lm88,ak88}).
This is called resonant spin-flavor (RSF) conversion.
The RSF conversions $\nu_e \leftrightarrow \bar{\nu}_{\mu,\tau}$
or $\bar{\nu}_{e} \leftrightarrow \nu_{\mu,\tau}$ have been investigated
in two-flavor models for solar (e.g., \cite{ak88,ak91,pu92,al93}) and 
SN (e.g., \cite{ab92,ap93,ts96,nq97}) neutrinos.
The RSF conversions of $\nu_e \leftrightarrow \bar{\nu}_e$
were also found \cite{ab92}.

SN neutrinos are most favorable to find evidence for RSF conversion.
Proto-neutron stars that have been formed at core collapse should have
strong magnetic field.
Observations of pulsars indicated that neutron stars have magnetic field
of the order of $10^{12}$ G.
Some pulsars indicate much stronger magnetic field with 
$\sim 10^{14} - 10^{15}$ G, so that they are called magnetars
(e.g., \cite{td95}).
The magnetic field in the outer region of Fe core of presupernova stars is
evaluated to reach $\sim 10^{10}$ G \cite{hw05}.
Owing to such a strong magnetic field and neutrino magnetic moment,
spin-flavor conversion of neutrinos is expected to occur.
Detailed features of the RSF conversions have been investigated
considering three flavors of neutrinos and antineutrinos numerically
\cite{as03a,as03b,as03c}
and analytically \cite{af03,am03}.
They have shown that very large flavor conversion can occur between
neutrinos and antineutrinos in SN ejecta.

Continuous efforts of numerical studies on SN explosions revealed
detailed conditions in the deep region of SN ejecta (e.g., \cite{jl07}).
One of the findings is the time evolution of electron mole fraction $Y_e$.
The electron fraction of the central core of a presupernova is 
between 0.4 and 0.5 \cite{hw01}.
After the core bounce in a scenario of delayed SN explosion, neutrino
heating helps the explosion of a SN.
The neutrino interaction also changes the electron fraction of the inner
region of the exploding materials \cite{aj07}.
The electron fraction in the neutrino-heated bubbles and 
the neutrino-driven winds exceeds 0.5 in several seconds.
This excess in $Y_e$ opens a new nucleosynthesis process, which is called
the $\nu p$ process \cite{pw05,ph06,fr06,fm06,wa06}.
The $\nu p$ process is one of the most promising processes to produce
proton-rich nuclei.
The electron fraction is expected to decrease eventually below 0.5 
in several seconds.
After electron fraction becomes small, $r$ process is expected to occur
in neutrino-driven winds (e.g., \cite{ww94,ot00,wi02}).

We expect that the change of the electron fraction also affects the
neutrino oscillations, especially RSF conversion.
The resonance densities of two RSF conversions depend on $1/|1-2Y_e|$, so that
the resonance densities will drastically change in accordance with 
the change in $Y_e$.
Furthermore, the relation between mass eigenstates and flavor eigenstates
will change at $Y_e = 0.5$.
Therefore, we expect that new neutrino signals will be detected in a 
SN explosion in our Galaxy, if neutrinos are Majorana particles
and have finite magnetic moment ($\sim 10^{-12} \mu_B$), and the SN
has strong magnetic field.
Although there are many studies on the RSF conversion 
(e.g., \cite{as03a,as03b,as03c,af03}), this type of the RSF conversion
effect on SN neutrino signal has not been studied very well.

In this study, we investigate detailed dependence of RSF conversion
effects of SN neutrinos on the electron fraction in
SN ejecta.
In particular, we pay attention to the difference of RSF conversion at the 
electron fraction  below and above 0.5.
We will discuss the time variation of a SN neutrino signal along
the evolutionary change in $Y_e$ and the dependence on neutrino oscillation 
parameters, i.e., mass hierarchy and the mixing angle $\theta_{13}$.

We organize this article as follows.
In Sec. II, the SN explosion model adopted in this study is described.
We adopted the time evolution of the density profile calculated using
the adiabatic explosion model in \cite{kk08}.
The energy spectra of the neutrinos and the neutrino oscillation
parameters are also written in this section.
In Sec. III we describe the relation of the effective squared mass 
with the density and the conversion probabilities due to the RSF conversion 
and the MSW effect.
Based on the obtained characteristics, the dependence of the neutrino
spectra and expected neutrino signals on the electron fraction,
magnetic field, and the neutrino oscillation parameters is shown.
In Sec. IV we discuss the time evolution of the SN neutrino
signal and constraints on the neutrino oscillation parameters and magnetic
moment.
We also discuss the relations of assumed magnetic field and the density
profile in the deep region of the SN ejecta.
Finally, in Sec. V we summarize this study.

\section{Model}

\subsection{Supernova explosion model}

\begin{figure}
\begin{center}
\includegraphics[width=8.5cm]{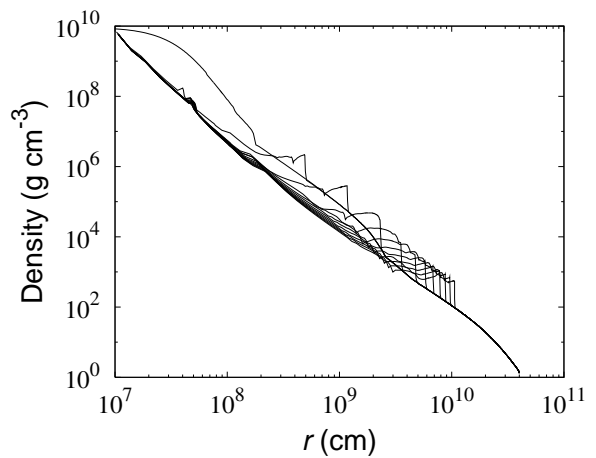}
\end{center}
\caption{\label{fig:sn-density} 
The density distribution of the SN ejecta at $t$ = 0, 0.5, 1, 2,
3, 4, 5, 6, 7, 8, 9, and 10 s.
The progenitor is 15 $M_\odot$ presupernova star \cite{ww95}.
The shock propagation is calculated in \cite{kk08}.
}
\end{figure}

In order to study the time evolution of the neutrino signals affected by
the RSF conversion and the MSW effect, we use the time evolution
of the density profile including the shock propagation in
the stellar interior used in \cite{kk08}.
The progenitor model is a 15 $M_\odot$ presupernova of \cite{ww95}.
The core collapse and shock propagation are calculated taking account of 
general relativity and assuming spherical symmetry.
The neutrino transport is not taken into account by assuming adiabatic
explosion.
Details of the hydrodynamical simulation are written in \cite{kk08}.

The snapshots of the density profile of the SN at 
$t = 0, 0.5, 1, 2, 3, 4, 5, 6, 7, 8, 9$, and 10 s are shown in Fig. 1.
A very wide range of the density from $10^{10}$ ${\rm g cm^{-3}}$ to 
1 ${\rm g cm^{-3}}$ is shown.
Therefore, we take account of the shock propagation effect on the RSF
conversion which will occur in a high density region.
On the other hand, this model assumes adiabatic explosion as mentioned above.
The evolution of electron fraction is not considered.
In this study, we assume that the electron fraction is constant 
in the innermost hot-bubble and wind region where the mass coordinate is 
smaller than 1.43 $M_\odot$.
We consider two cases of the electron fraction, $Y_e$ = 0.49 and 0.51.
We also consider the simple time variation model for $Y_e$ in this discussion.
The electron fraction of the outer ejecta is assumed to be the
same as that of the progenitor model because the electron fraction
does not change by the neutrino irradiation there.

We note that the electron fraction inside the proto-neutron star should be 
very small due to neutronization.
However, the density in the region is much larger than the resonance density
of the RSF-H conversion, so that virtually no conversion is expected there.
Thus, we neglect the change of the electron fraction in this region.

The magnetic field profile of the SN is quite uncertain.
In this study, we assume that the SN magnetic field transverse to
the neutrino propagation changes as $\propto r^{-3}$ according to 
\cite{as03a,as03b,as03c}.
The magnetic field $B_\bot$ is described as 
\begin{equation}
B_\bot = B_0 \left( \frac{r_0}{r} \right)^3,
\end{equation}
where $r_0 = 10^8$ cm and $B_0$ is the magnetic field at $r_0$.
The value of $B_0$ is used as a parameter.
Typical strength is set to be $B_0 = 10^{11}$ G.
We discuss the magnetic field in massive stars and SNe in Sec. IV.

\subsection{Supernova neutrino model}

We use the SN neutrino model adopted in \cite{yk06a,yk06b,ys08}.
The neutrinos are assumed to carry the energy of 
$E_{\nu,total} = 3 \times 10^{53}$ erg \cite{wh90}.
The neutrino luminosity is equally partitioned among three flavors of
neutrinos and antineutrinos.
It is assumed to decay exponentially with the decay time of 3 s
\cite{yk06a,yk06b,ys08,wh90,yk05}.
The energy spectra of neutrinos emitted from a proto-neutron star 
are assumed to obey Fermi-Dirac distributions with zero chemical potential.
The temperatures of $\nu_e$, $\bar{\nu}_e$, and $\nu_x$, where $\nu_x$
correspond to $\nu_{\mu,\tau}$ and $\bar{\nu}_{\mu,\tau}$, are
set to be $(T_{\nu_e}, T_{\bar{\nu}_e}, T_{\nu_x})$ = 
(3.2, 5, 6) MeV \cite{yk06a,yk06b,ys08}.

Neutrinos change their flavors along the propagation in the stellar interior.
We consider three flavor neutrino oscillations taking account of
the MSW effect and the RSF conversion.
The equation of the neutrino flavor change is written as
(e.g. \cite{as03a,af03})
\begin{equation}
i \frac{d}{dr}
\left( \begin{array}{c}
\nu \\ \bar{\nu}
\end{array} \right)
=\left\{ \left( \begin{array}{cc}
H & MB_\bot \\ -MB_\bot & H \end{array} \right)
+ \left( \begin{array}{cc}
V & 0 \\ 0 & \bar{V} \end{array} \right) \right\}
\left( \begin{array}{c} \nu \\ \bar{\nu} \end{array} \right)
\end{equation}
where
\begin{equation}
\nu = \left(\begin{array}{c}
\nu_e \\ \nu_\mu \\ \nu_\tau \end{array} \right) , \quad
\bar{\nu} = \left(\begin{array}{c}
\bar{\nu}_e \\ \bar{\nu}_\mu \\ \bar{\nu}_\tau \end{array} \right) , 
\end{equation}
\begin{equation}
H = U \left(\begin{array}{ccc}
0 & 0 & 0 \\ 
0 & \frac{\Delta m^2_{21}}{2 E_\nu} & 0 \\
0 & 0 & \frac{\Delta m^2_{31}}{2 E_\nu} 
\end{array} \right) U^\dagger ,
\end{equation}
\begin{equation}
M = \left(\begin{array}{ccc}
0 & \mu_{e \mu} & \mu_{e \tau} \\ 
-\mu_{e \mu} & 0 & \mu_{\mu \tau} \\
-\mu_{e \tau} & -\mu_{\mu \tau} & 0
\end{array} \right) ,
\end{equation}
\begin{eqnarray}
V = -\bar{V} &=& \left(\begin{array}{ccc}
V_{\nu_e} & 0 & 0 \\ 0 & V_{\nu_\mu} & 0 \\ 0 & 0 & V_{\nu_\tau}
\end{array} \right) \\
&=& \frac{\sqrt{2}}{2} G_F \frac{\rho}{m_u}
\left(\begin{array}{ccc}
3Y_e-1 & 0 & 0 \\ 0 & Y_e-1 & 0 \\ 0 & 0 & Y_e-1 \end{array} \right) ,
\nonumber
\end{eqnarray}
\begin{equation}
U = \left(\begin{array}{ccc}
c_{12}c_{13} & s_{12}c_{13} & s_{13} \\
-s_{12}c_{23}-c_{12}s_{23}s_{13} & c_{12}c_{23}-s_{12}s_{23}s_{13} &
s_{23}c_{13} \\
s_{12}s_{23}-c_{12}c_{23}s_{13} & -c_{12}s_{23}-s_{12}c_{23}s_{13} &
c_{23}c_{13} \end{array} \right) ,
\end{equation}
$\Delta m^2_{ij} = m^2_i - m^2_j$, $m_i$ is the mass of the $i$th mass
eigenstate neutrinos, $E_\nu$ is the
neutrino energy, $G_F$ is the Fermi constant, $\rho$ is the density,
$m_u$ is the atomic mass unit, $\mu_{\alpha \beta}$ is the transition
neutrino magnetic moment, $B_\bot$ is the transverse magnetic field,
$s_{ij} = \sin \theta_{ij}$, and $c_{ij} = \cos \theta_{ij}$.
Here we use the units of $\hbar = c = 1$.

As shown in \cite{af03}, there are two MSW resonances and three RSF 
resonances. 
The resonance densities are written as
\begin{eqnarray}
\rho_{res}({\rm MSW-H}) &=&
\frac{m_u |\Delta m^2_{31}| \cos2\theta_{13}}{2\sqrt{2} G_F E_\nu}
\frac{1}{Y_e} , \\
\rho_{res}({\rm MSW-L}) &=&
\frac{m_u \Delta m^2_{21} \cos2\theta_{12}}{2\sqrt{2} G_F E_\nu}
\frac{1}{Y_e} , \\
\rho_{res}({\rm RSF-H}) &=&
\frac{m_u |\Delta m^2_{31}| \cos2\theta_{13}}{2\sqrt{2} G_F E_\nu}
\frac{1}{|1-2Y_e|} , \\
\rho_{res}({\rm RSF-L}) &=&
\frac{m_u \Delta m^2_{21} \cos2\theta_{12}}{2\sqrt{2} G_F E_\nu}
\frac{1}{|1-2Y_e|} , \\
\rho_{res}({\rm RSF-X}) &=&
\frac{m_u |\Delta m^2_{31}| \cos^2\theta_{13}}{2\sqrt{2} G_F E_\nu}
\frac{1}{1-Y_e} ,
\end{eqnarray}
where suffixes ^^ ^^ H'' and ^^ ^^ L'' correspond to the high and low resonance
densities deduced using $|\Delta m^2_{31}|$ and $\Delta m^2_{21}$,
respectively.
When $\sin^22\theta_{13}$ is larger than a critical value, the RSF-X 
resonance changes to the RSF-E resonance \cite{af03}.
The RSF-E resonance appears in place of RSF-X resonance owing to a nonlinear
dependence of the effective squared masses on the density.
We do not distinguish the RSF-E resonance from the RSF-X and use 
a common term, the RSF-X resonance, in this study.

Most of neutrino oscillation parameters have been determined precisely
by recent neutrino experiments \cite{sno04,sk06,kl08}.
We use the squared mass differences $\Delta m^2_{ij}$ as follows:
\begin{eqnarray}
\Delta m^2_{21} &=& 7.6 \times 10^{-5} {\rm eV^2} \quad {\rm and} \\
|\Delta m^2_{31}| &=& 2.5 \times 10^{-3} {\rm eV^2}. \nonumber
\end{eqnarray}
The mixing angles are taken as
\begin{equation}
\sin^22\theta_{12} = 0.87 \quad {\rm and} \quad
\sin^22\theta_{23} = 1.
\end{equation}
The mass hierarchy, i.e., the sign of $\Delta m^2_{31}$, and the mixing
angle $\sin^22\theta_{13}$ are taken as parameters.
We consider both of normal and inverted mass hierarchies.
For the mixing angle $\sin^22\theta_{13}$, we consider 
$\sin^22\theta_{13} = 0.04$ (adiabatic MSW-H case) and $1 \times 10^{-6}$
(nonadiabatic MSW-H case).
The former value corresponds to adiabatic conversion at the MSW-H
resonance in the presupernova density profile.
The latter value corresponds to nonadiabatic at the MSW-H resonance.

We assume that the transition magnetic moment of neutrinos 
$\mu_{\alpha \beta}$ does not depend on flavors.
We set $\mu_{\alpha \beta} = 1 \times 10^{-12} \mu_B$ commonly.
This value is the same as that adopted in \cite{as03a,as03b,as03c}.
Since the effect of neutrino magnetic moment appears in the form of 
$\mu_{\alpha \beta}B$, small magnetic moment with strong magnetic field may
bring about a similar effect to the case of large magnetic moment
with weak magnetic field.

\subsection{Detection of neutrino signals}

We evaluate the event rate of the neutrino signals emitted from a SN
using a water-{\v C}erenkov detector of Super-Kamiokande (SK).
The positron energy spectrum of the rate of neutrino event $i$ is 
evaluated to be
\begin{eqnarray}
\frac{d^2N_i}{dE_e dt} &=& \nonumber
\frac{N_{{\rm target},i}}{4 \pi d^2} \eta({E_e}) \\ \nonumber
&\times& \sum_{\beta} \int_{0}^{\infty}
\frac{d^2N_{\nu_\beta}}{dE_\nu dt}
P_{\nu_\beta \rightarrow \nu_\alpha}(E_\nu)
\sigma_{\nu_{\alpha} i}(E_\nu) \\
&\times& \left( \frac{dE_\nu}{dE_e'}\right)_i
R(E_e, E_e') dE_e' ,
\end{eqnarray}
where $N_{{\rm target},i}$ is the number of the target for reaction $i$,
$d$ is the distance from a SN, $\eta(E_e)$ is the 
detection efficiency, $d^2N_{\nu_\beta}/dE_\nu dt$
is the $\nu_\beta$ number rate per neutrino energy per time, 
$P_{\nu_\beta \rightarrow \nu_\alpha}(E_\nu)$ is the transition
probability from $\nu_\beta$ to $\nu_\alpha$ with the energy 
$E_\nu$ through neutrino oscillations, 
$\sigma_{\nu_\alpha i}(E_\nu)$ is the cross section of $\nu_\alpha$
neutrino reaction $i$, 
$(dE_\nu/dE_e')_i$ is the derivative of the incident 
neutrino energy on the emitted positron energy for 
reaction $i$, $R(E_e, E_e')$ is the energy resolution 
function in the Gaussian form with the width of 
$\Delta(E_e)=0.2468+0.1492\sqrt{E_e}+0.0690E_e$
MeV \cite{sk08}.
The rate of the neutrino event $i$ is evaluated as
\begin{equation}
\frac{dN_i}{dt} = \int_{E_{th}}^{\infty} 
\frac{d^2N_i}{dE_e dt} dE_e,
\end{equation}
where $E_{th}$ is the lowest detectable energy.
We suppose a water-{\v C}erenkov detector of 22.5 kton fiducial volume
corresponding to SK detector.
We consider that the distance from a SN is 10 kpc, which is close to the
distance from the Galactic center.
The event rate for other distances is easily transformed using the relation
$d^2N_i/dE_e dt \propto d^{-2}$.
We assumed that the detection efficiency is unity when $E_e$ is
larger than $E_{th} = 5$ MeV and otherwise zero supposing
the third phase of the SK experiment (SK-III) \cite{ta08}.
We consider the antineutrino-proton reaction
\begin{equation}
p + \bar{\nu}_e \rightarrow n + e^+.
\end{equation}
The reaction rate is adopted from \cite{sv03}.

\section{Results}
\subsection{Effective squared masses}

\begin{figure*}
\includegraphics[angle=-90,width=6.7cm]{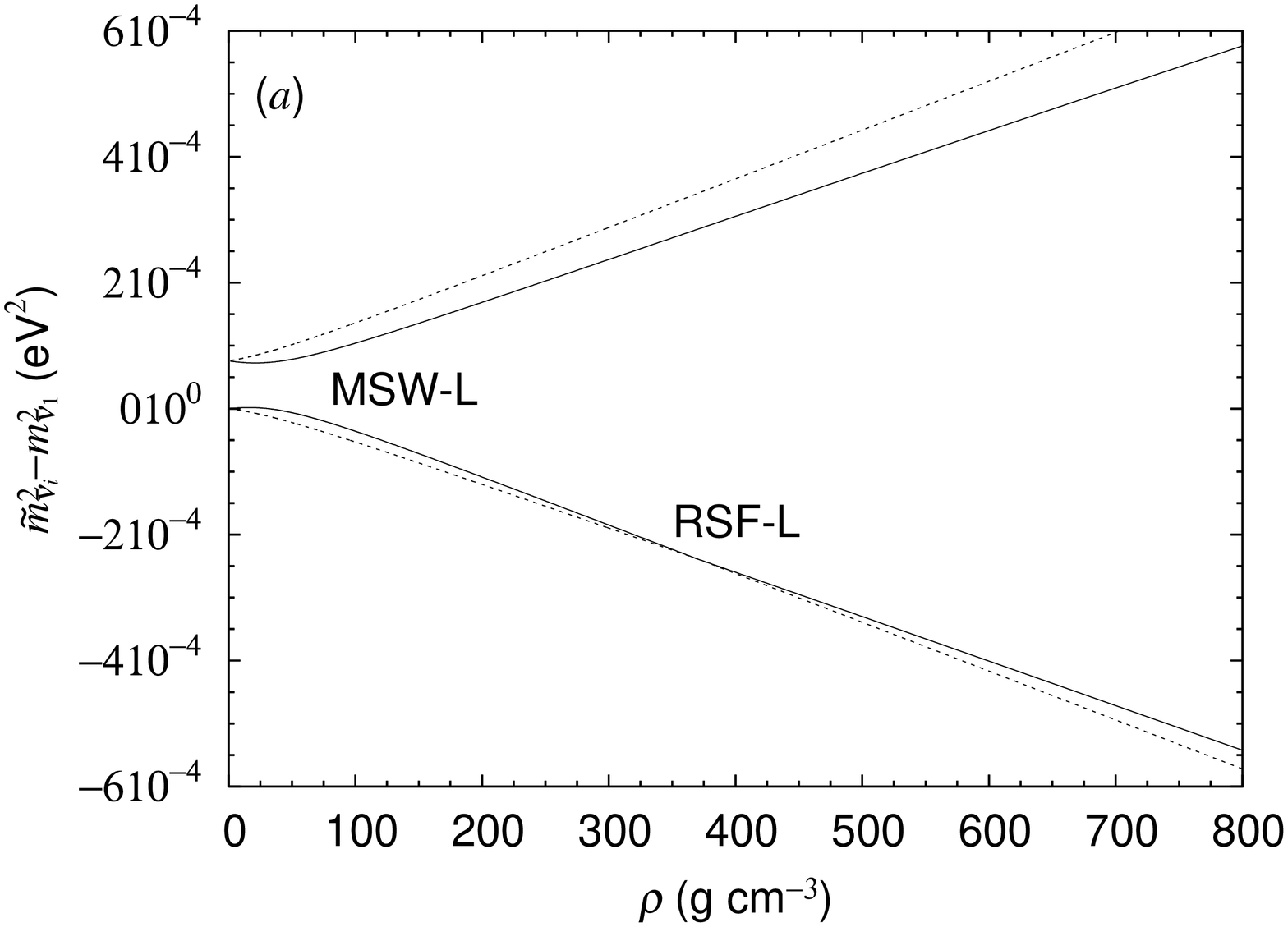}
\includegraphics[angle=-90,width=6.7cm]{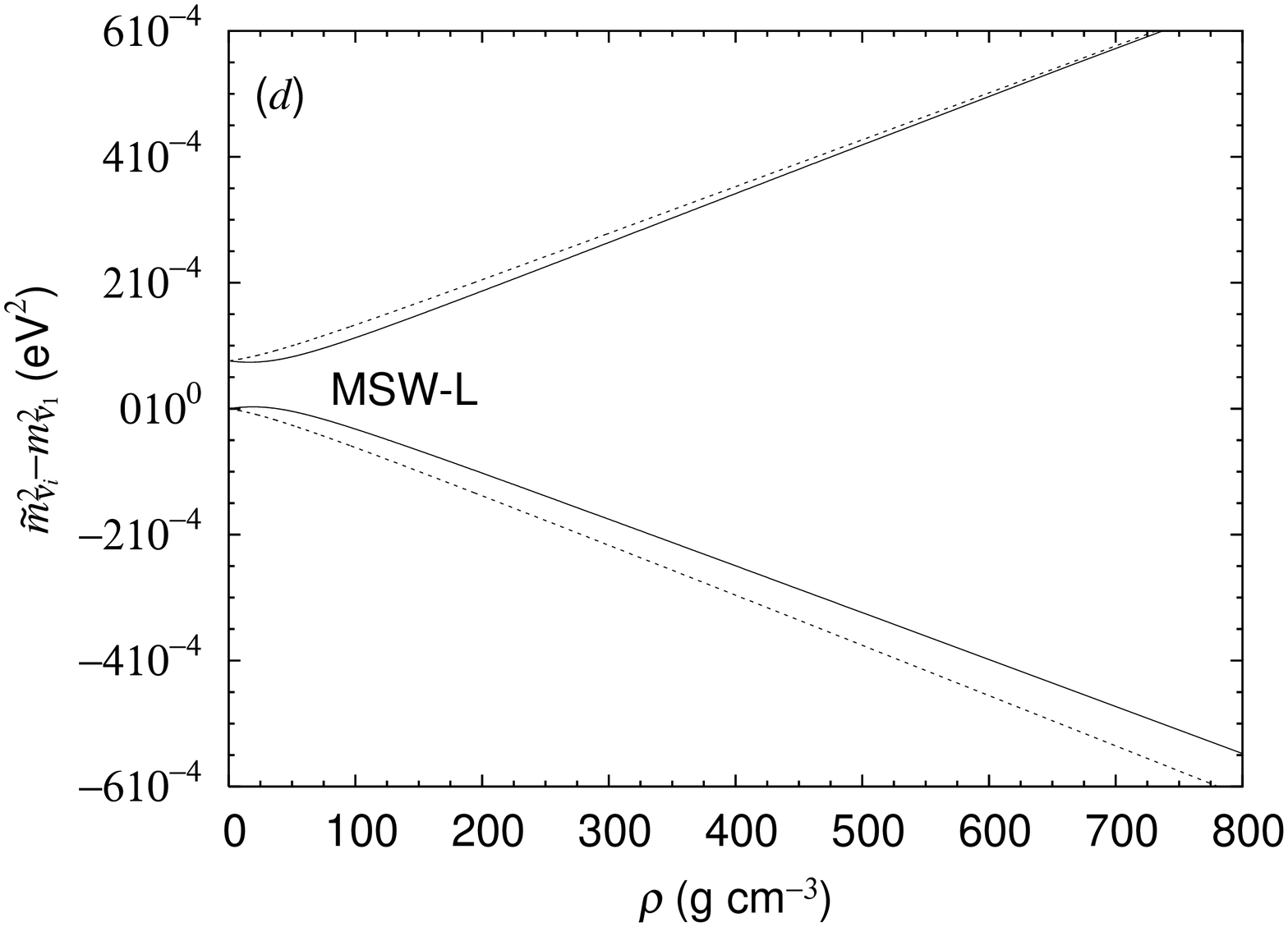}
\includegraphics[angle=-90,width=6.7cm]{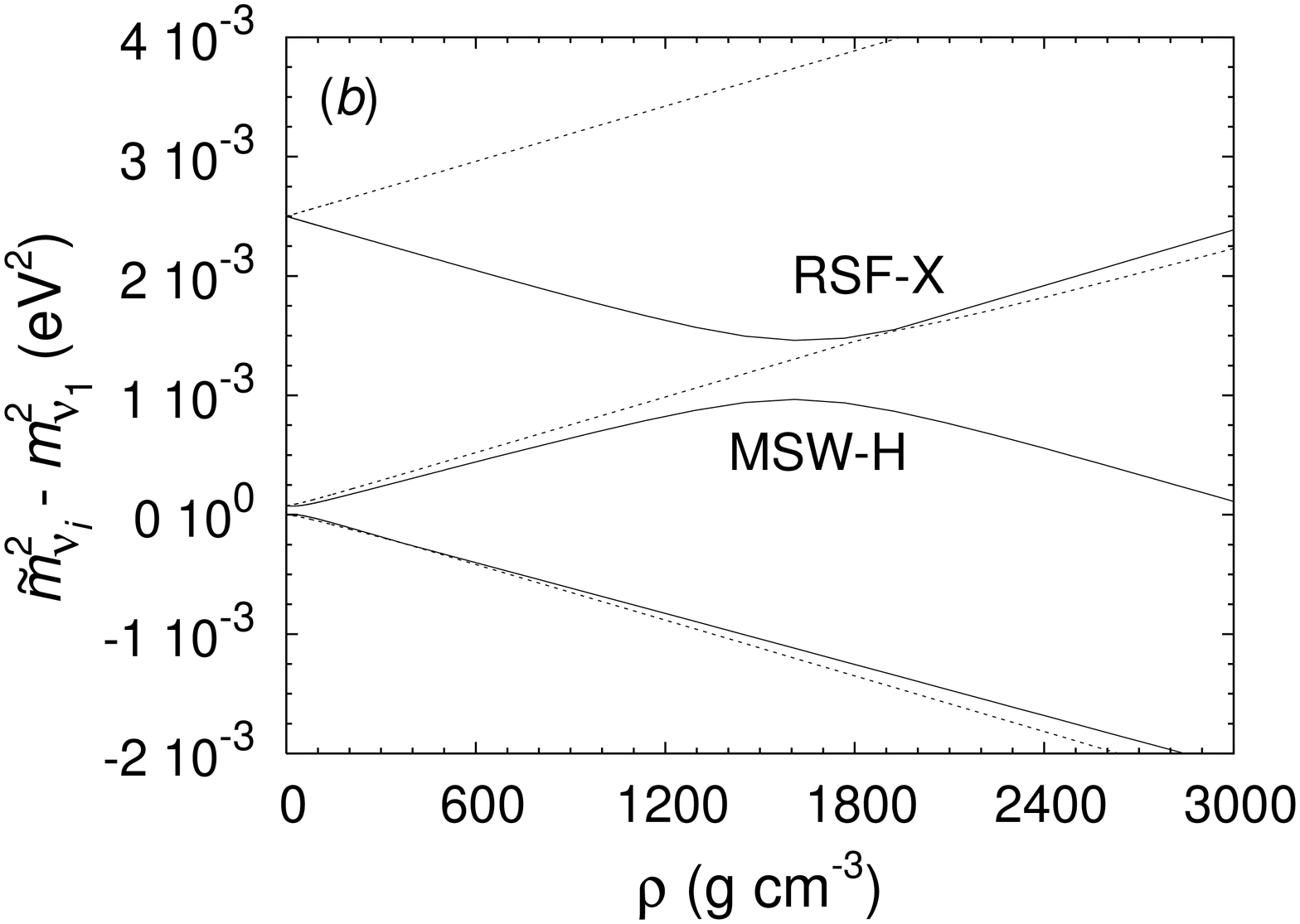}
\includegraphics[angle=-90,width=6.7cm]{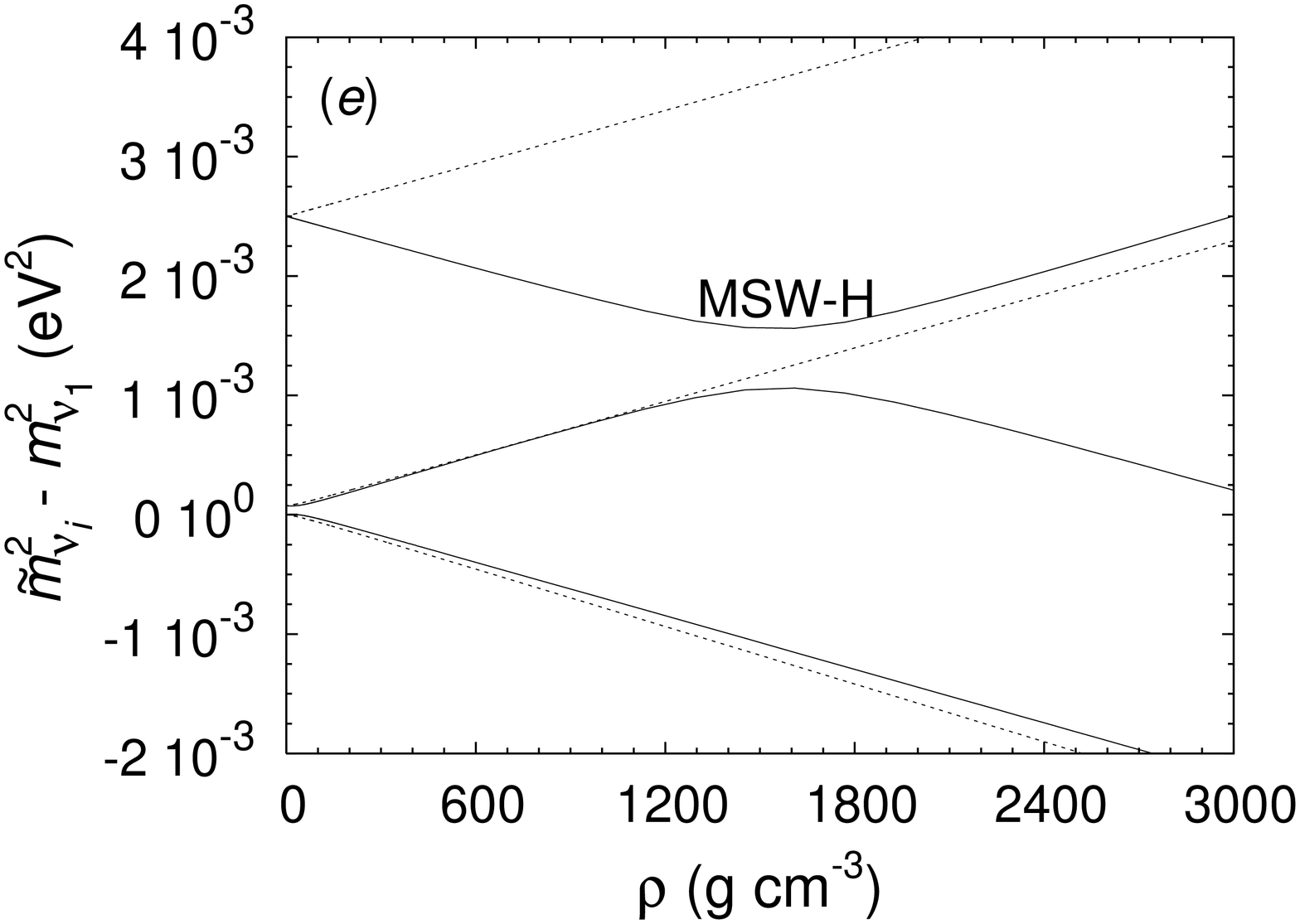}
\includegraphics[angle=-90,width=6.7cm]{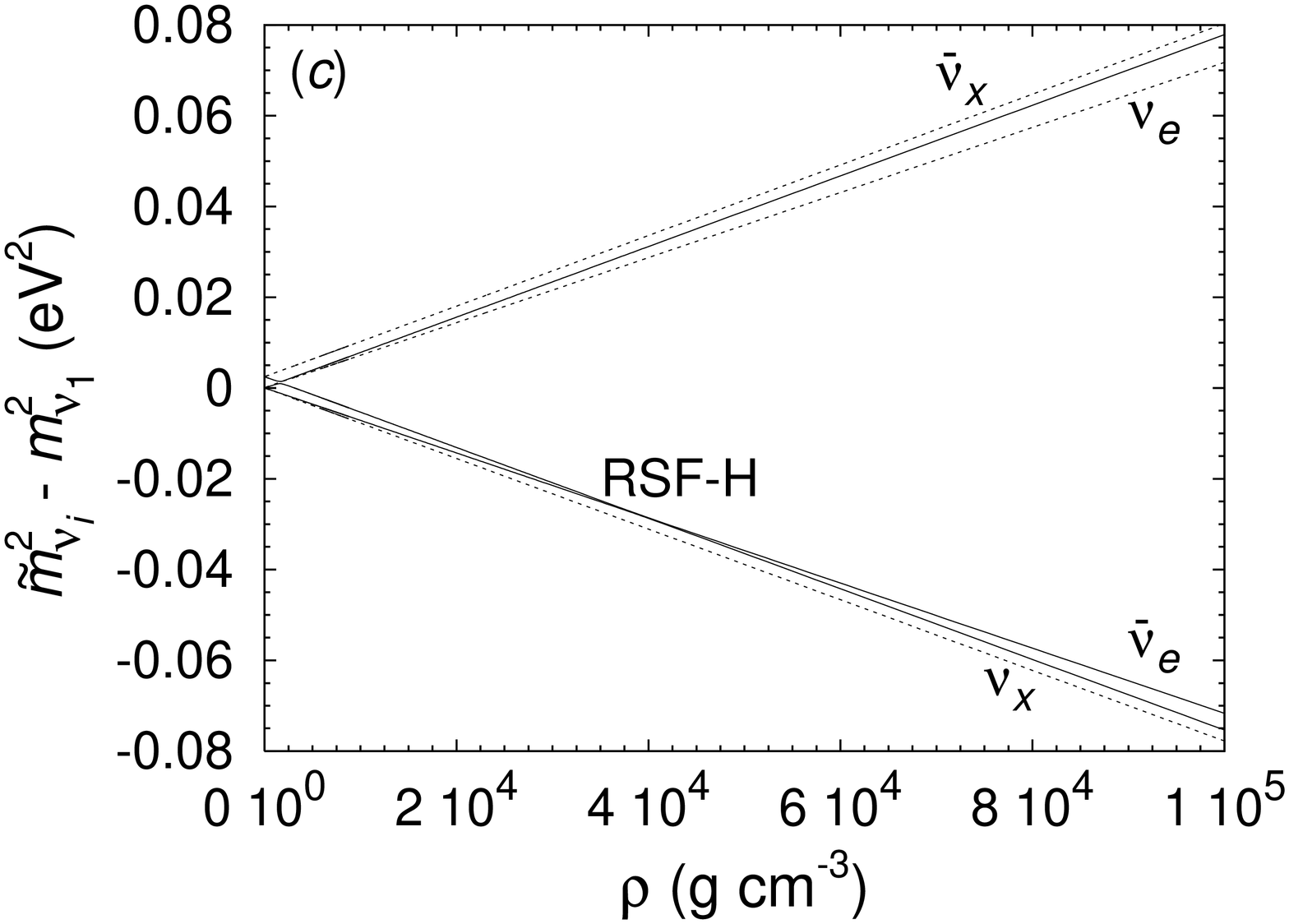}
\includegraphics[angle=-90,width=6.7cm]{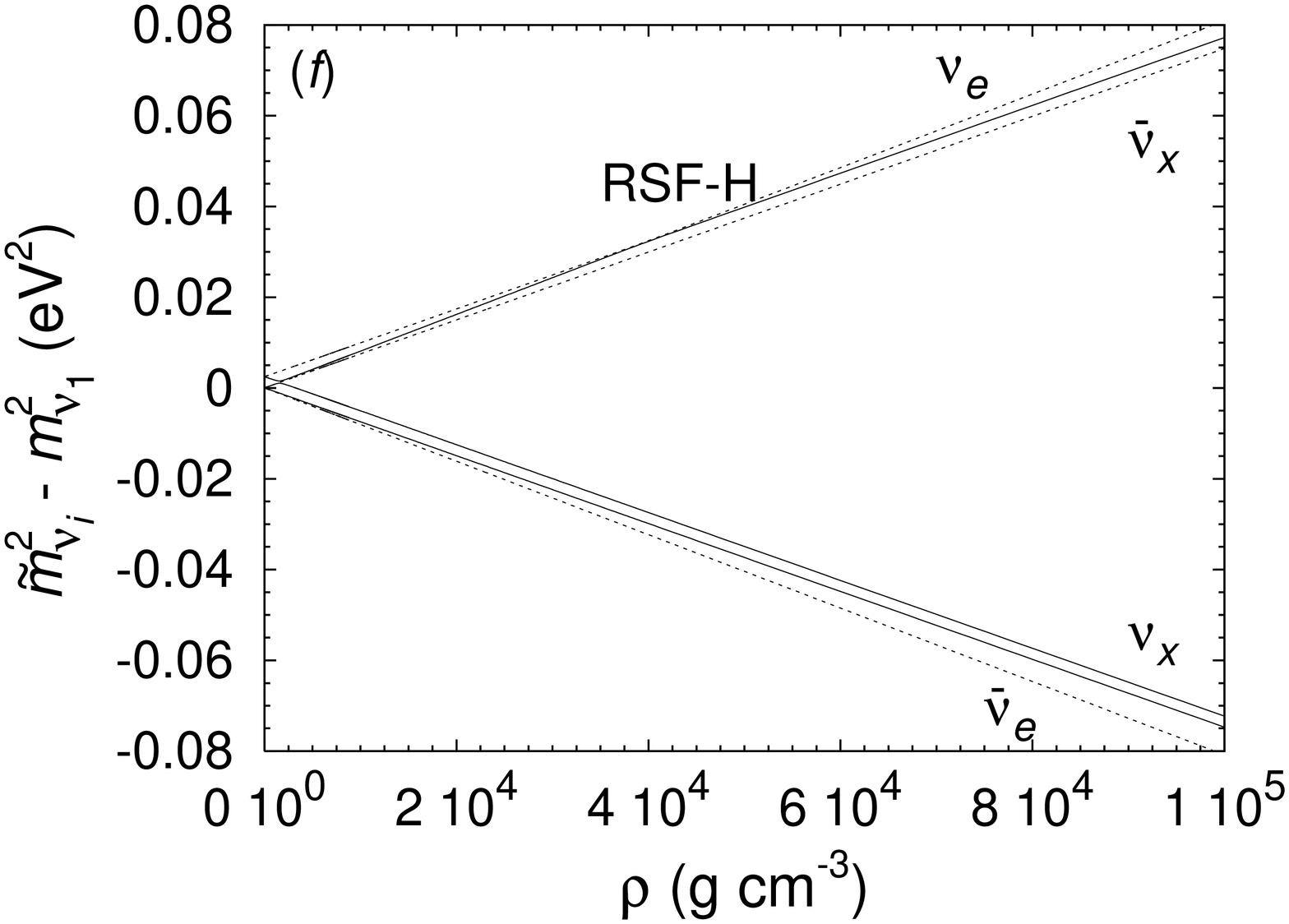}
\caption{\label{fig:mfnorm} 
Effective squared masses of mass eigenstates with the relation to the density
in the case of normal mass hierarchy.
Electron fraction is assumed to be 0.49 for (a)$-$(c),
and 0.51 for panels (d)$-$(f).
Solid lines and dotted lines show the effective squared masses of neutrinos
and antineutrinos, respectively.
For each line type, mass eigenstates 1, 2, and 3 correspond in ascending
order of squared mass.
At the RSF-H and MSW-H resonances, ($\rho_{res}$, $r_{res}$) = 
($3.9 \times 10^{4}$ g cm$^{-3}$, $8.0 \times 10^8$ cm) and
($1.6 \times 10^{3}$ g cm$^{-3}$, $4.8 \times 10^9$ cm), respectively,
in the cases of $Y_e$ = 0.49 and 0.51 commonly.
}
\end{figure*}

It should be useful to know the dependence of effective squared masses
on the density when one evaluates the conversion probabilities of
neutrinos and antineutrinos through the RSF and MSW resonances.
We will compare the results between the cases of $Y_e = 0.49$ and 0.51.
We used the density profile of the 15 $M_\odot$ presupernova and the
magnetic field of $B_0 = 1 \times 10^{10}$ G.
The mixing angle $\sin^22\theta_{13}$ is assumed to be 0.04,
corresponding to the adiabatic MSW-H resonance.
The effective squared masses are different between neutrinos and antineutrinos 
in a finite density.
The relation of the effective squared masses is as follows:
$\tilde{m}^2_{\bar{\nu}_1} \le \tilde{m}^2_{\nu_1}$,
$\tilde{m}^2_{\nu_2} \le \tilde{m}^2_{\bar{\nu}_2}$, and
$\tilde{m}^2_{\nu_3} \le \tilde{m}^2_{\bar{\nu}_3}$, independent of 
mass hierarchy and the equalities are satisfied in vacuum.

Figure 2 shows the density dependence of the effective squared masses 
in normal mass hierarchy.
In the case of $Y_e = 0.49$, the dependence of
the squared masses is well explained in \cite{af03}.
We see five resonances, the MSW-H, L, and the RSF-H, L, and X.
In the high density limit, $\nu_e$ and $\bar{\nu}_e$ correspond to 
$\bar{\nu}_2$ and $\nu_2$, respectively.
At the same time, both $\nu_{\mu}$ and $\nu_{\tau}$ are the mixed states of
$\nu_1$ and $\bar{\nu}_1$.
Therefore, the conversion of $\bar{\nu}_e$ and $\nu_{\mu,\tau}$ occurs
at the RSF-H resonance.
The conversion of $\nu_e$ and $\bar{\nu}_{\mu,\tau}$ occurs at 
the RSF-X resonance.

In the case of the normal mass hierarchy and $Y_e = 0.51$, the dependence
of the effective squared masses is different from the case of $Y_e = 0.49$.
The effective squared mass of $\nu_e$ is the largest at high density limit.
That of $\bar{\nu}_e$ is the smallest.
The flavor conversion between $\nu_e$ ($\bar{\nu}_3$) and 
$\bar{\nu}_{\mu,\tau}$ ($\nu_3$) occurs at RSF-H resonance.
We see MSW-H and MSW-L resonances similar to the case of $Y_e = 0.49$.
There is no resonance corresponding to RSF-L or RSF-X resonance.
This is due to a nonlinear dependence of the squared masses on the density.

The correspondence between flavor and mass eigenstates at the high density
limit is recognized using the effective potentials.
The effective potential of $\nu_e$ is
\begin{equation}
V_{\nu_e} = \frac{\sqrt{2}}{2} G_F \frac{\rho}{m_u} (3 Y_e -1) .
\end{equation}
That of $\bar{\nu}_{\mu,\tau}$ is 
\begin{equation}
V_{\bar{\nu}_{\mu,\tau}} = \frac{\sqrt{2}}{2} G_F \frac{\rho}{m_u} (1 -Y_e) .
\end{equation}
The satisfying inequalities hold: $V_{\nu_e} < V_{\bar{\nu}_{\mu,\tau}}$ for
$Y_e < 0.5$ and $V_{\nu_e} > V_{\bar{\nu}_{\mu,\tau}}$ for $Y_e > 0.5$.
Therefore, the effective squared mass of $\nu_e$ is the largest at the high
density limit in $Y_e > 0.5$.
In the case of $Y_e = 0.5$, the equality 
$V_{\nu_e} = V_{\bar{\nu}_{\mu,\tau}}$ is satisfied and there are no
RSF-H and RSF-L resonances.

\begin{figure*}
\includegraphics[angle=-90,width=6.7cm]{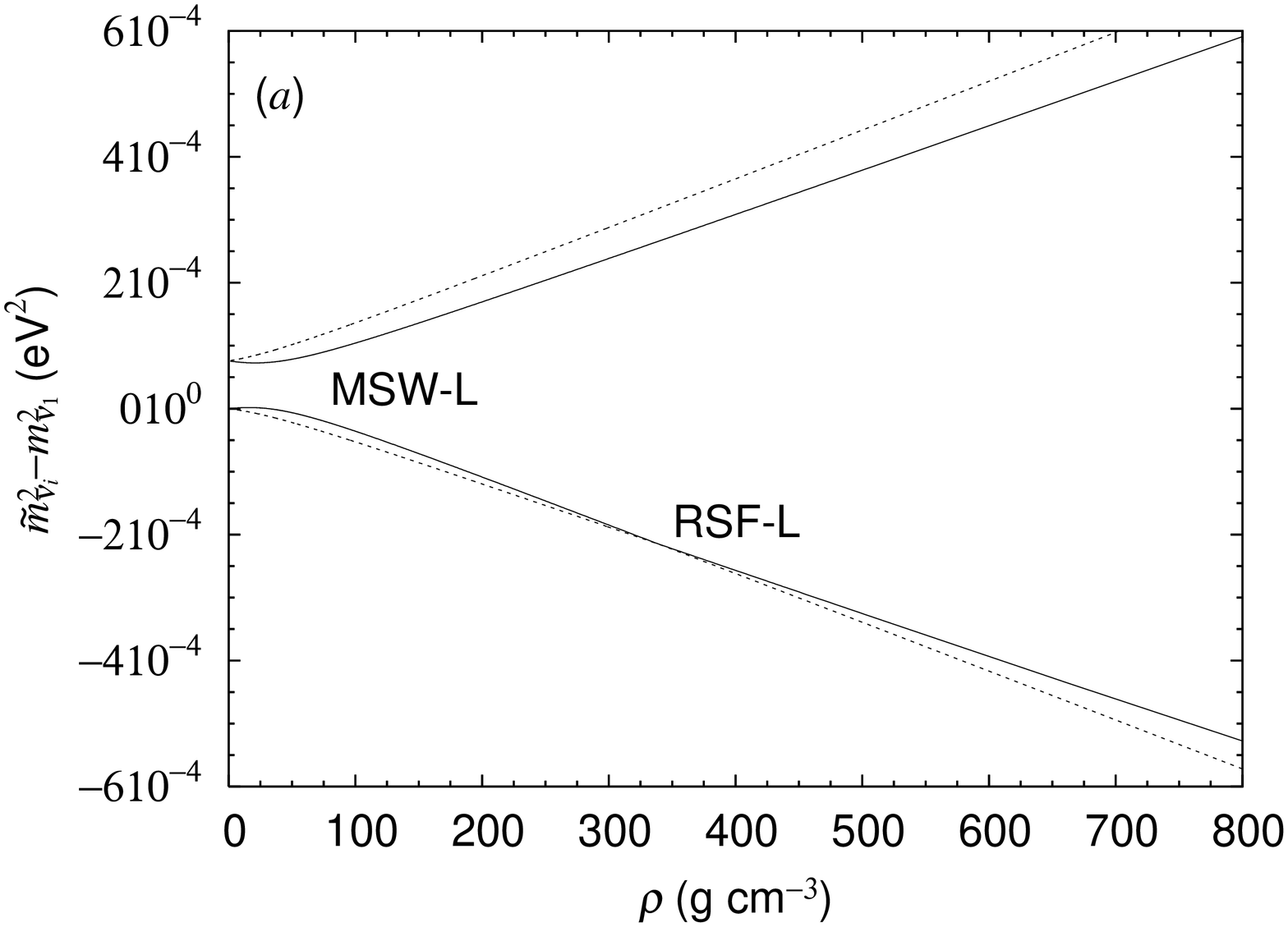}
\includegraphics[angle=-90,width=6.7cm]{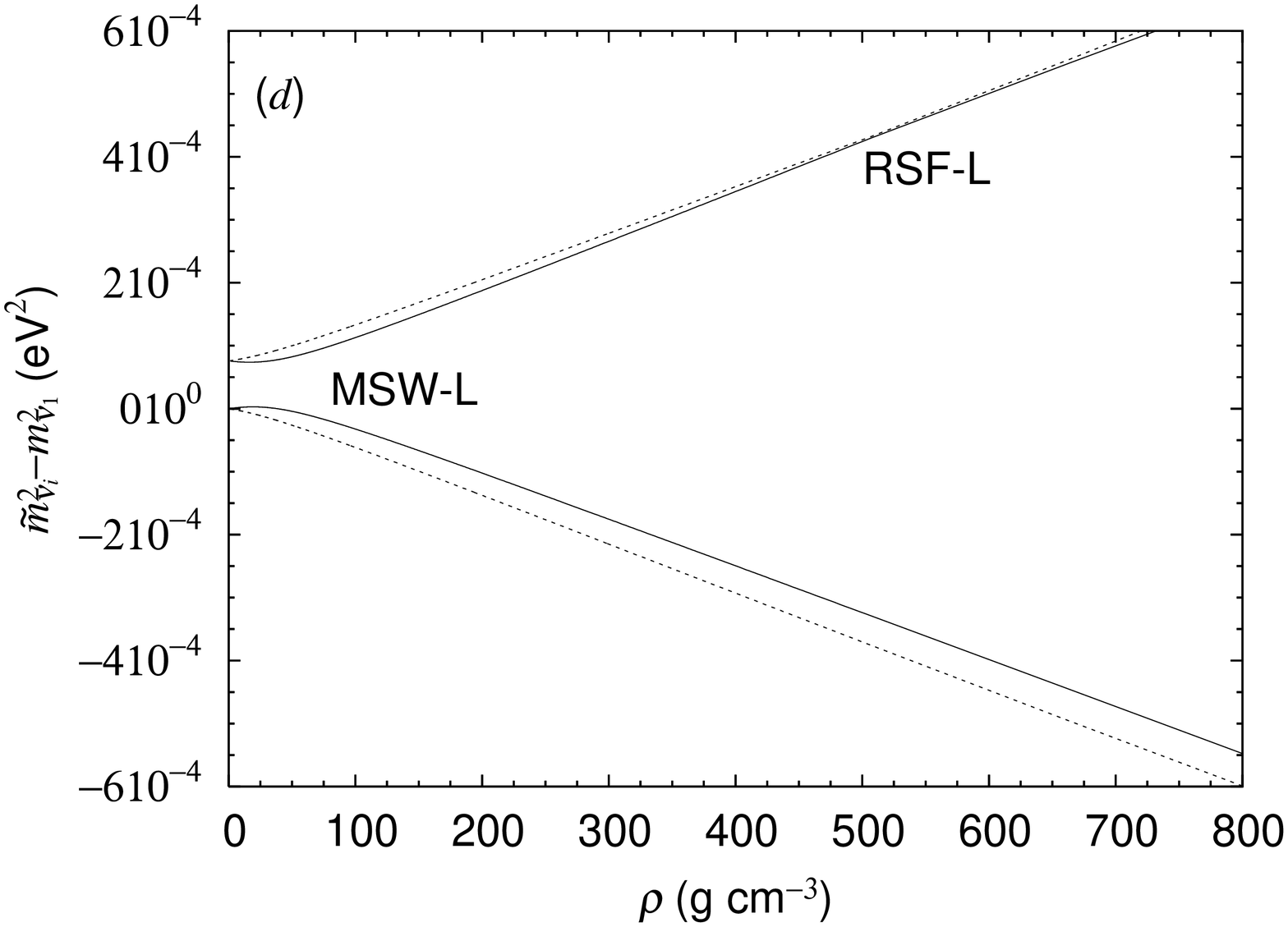}
\includegraphics[angle=-90,width=6.7cm]{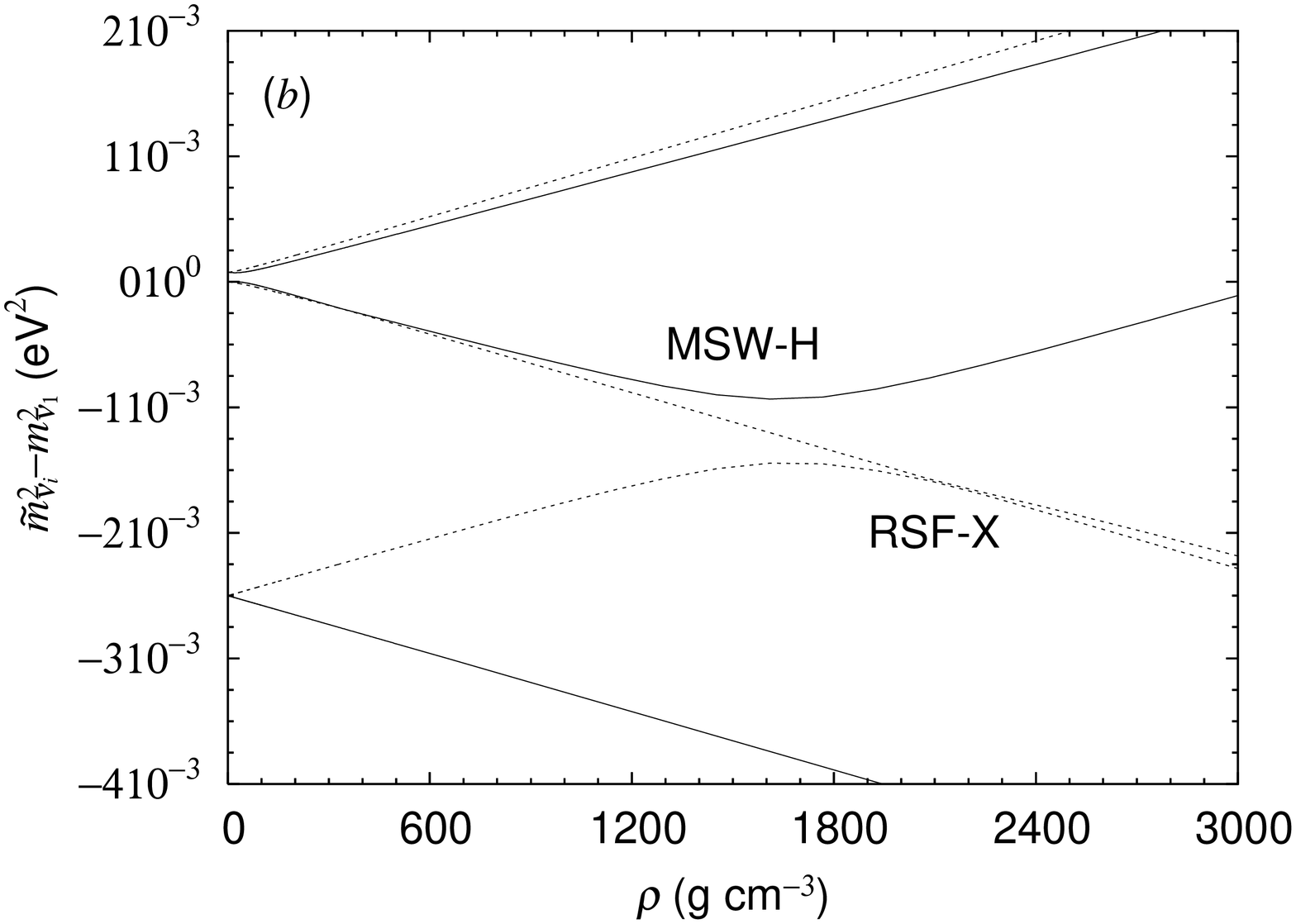}
\includegraphics[angle=-90,width=6.7cm]{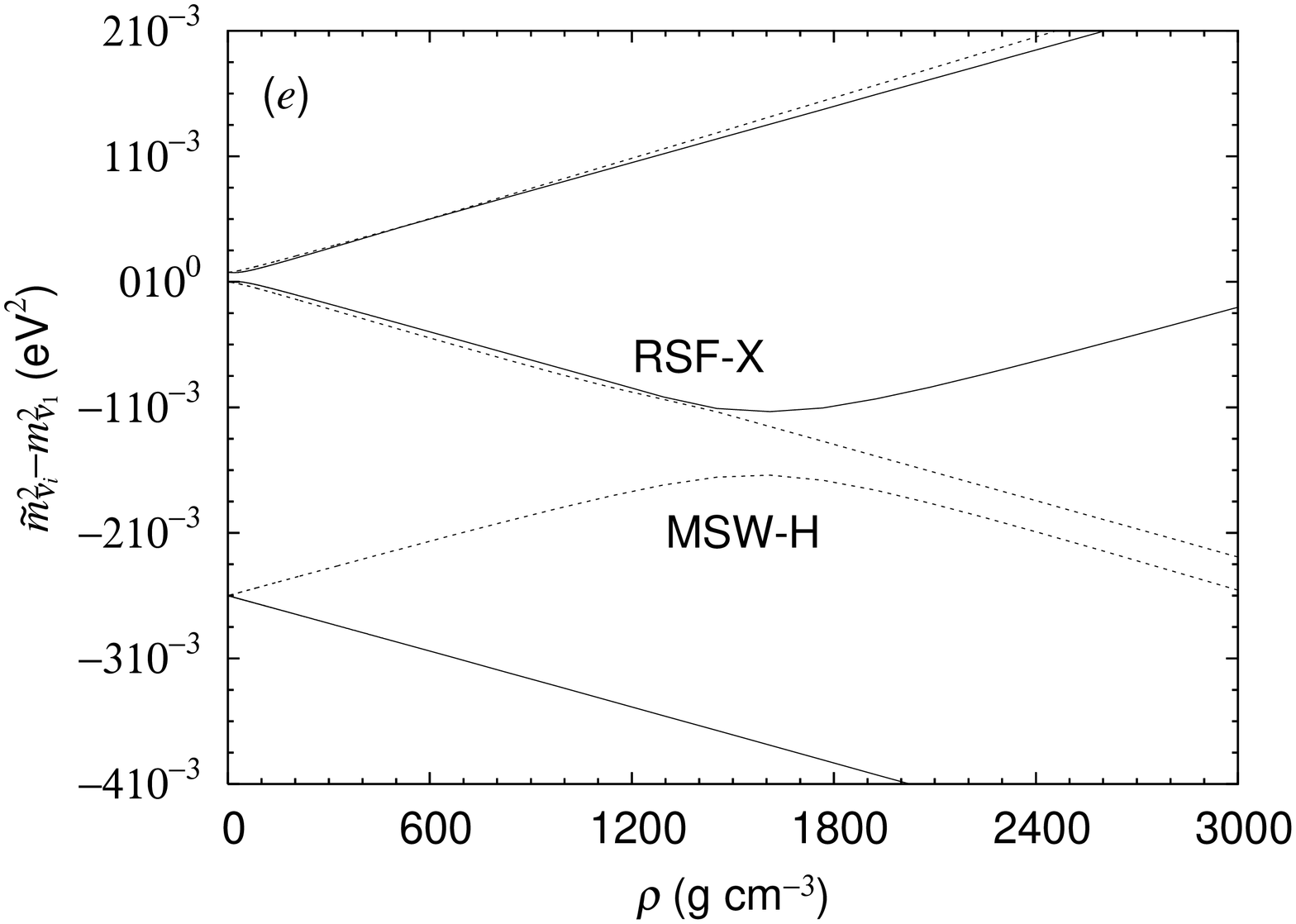}
\includegraphics[angle=-90,width=6.7cm]{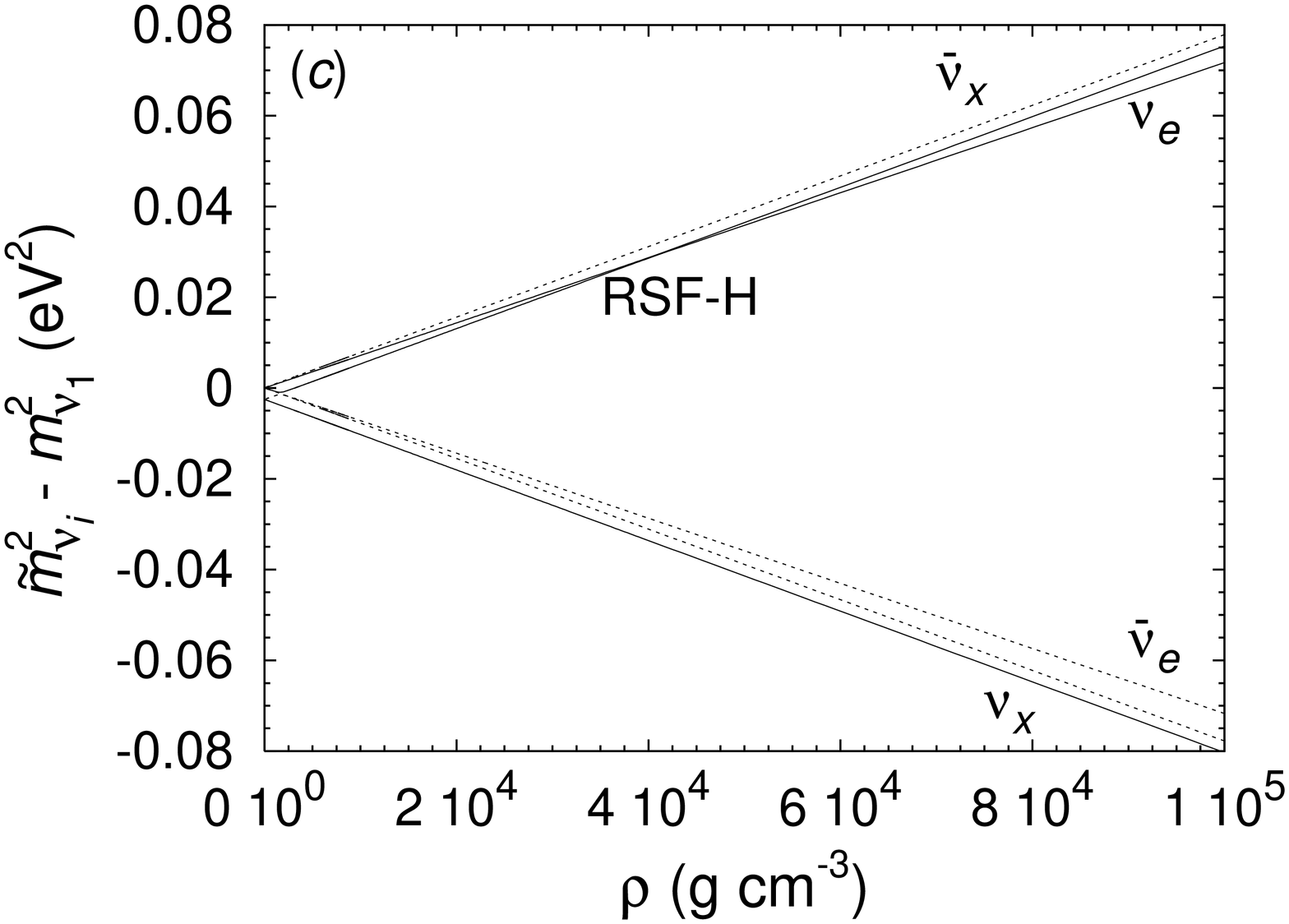}
\includegraphics[angle=-90,width=6.7cm]{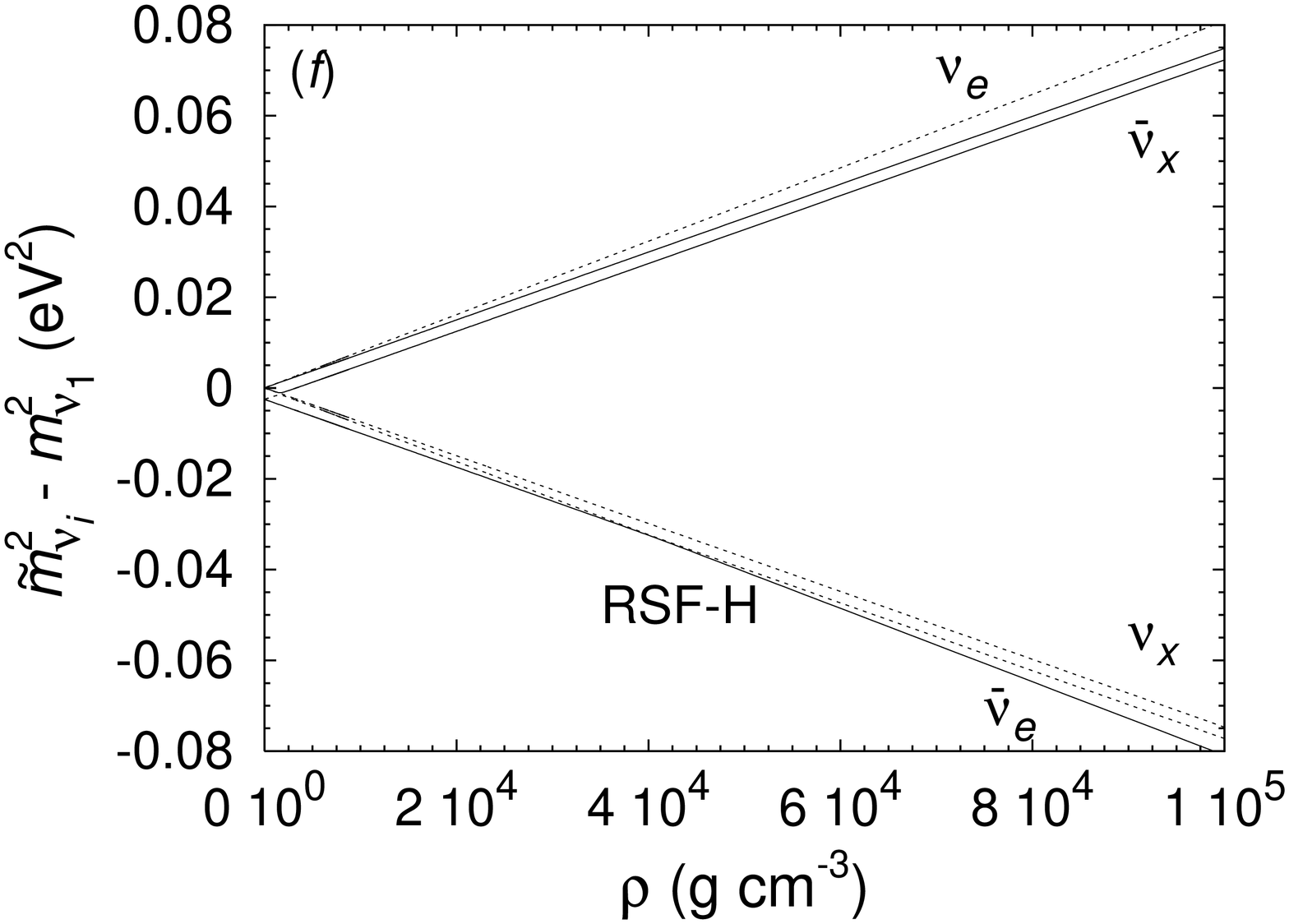}
\caption{\label{fig:mfinv}
Effective squared masses of mass eigenstates with the relation to the density
in the case of inverted mass hierarchy.
Electron fraction is assumed to be 0.49 for panels (a)$-$(c),
and 0.51 for panels (d)$-$(f).
Solid lines and dotted lines show the effective squared masses of neutrinos
and antineutrinos, respectively.
For each line type, mass eigenstates 3, 1, and 2 correspond in ascending
order of squared mass.
At the RSF-H and MSW-H resonances, ($\rho_{res}$, $r_{res}$) = 
($4.1 \times 10^{4}$ g cm$^{-3}$, $7.8 \times 10^8$ cm) and
($1.5 \times 10^{3}$ g cm$^{-3}$, $4.8 \times 10^9$ cm), respectively,
in the cases of $Y_e$ = 0.49 and 0.51 commonly.
}
\end{figure*}

Figure 3 shows the density dependence of the effective squared masses 
in the inverted mass hierarchy.
In the case of $Y_e = 0.49$, $\nu_e$ and $\bar{\nu}_e$ correspond to 
the mass eigenstates $\nu_1$ and $\bar{\nu}_1$, respectively, at high
density limit.
Flavor eigenstates $\nu_{\mu,\tau}$ and $\bar{\nu}_{\mu,\tau}$ are the
mixed states of $\nu_2$ and $\bar{\nu}_2$, and $\nu_3$ and $\bar{\nu}_3$,
respectively.
There are the RSF-H, RSF-L, and RSF-X resonances as seen in 
the case of the normal mass hierarchy.
The conversion between $\nu_e$ and $\bar{\nu}_{\mu,\tau}$ occurs at the RSF-H 
resonance.
The conversion between $\bar{\nu}_e$ and $\nu_{\mu,\tau}$ occurs at the RSF-L 
resonance.
The RSF-X resonance converts $\bar{\nu}_1$ and $\bar{\nu}_3$.
The converting flavors depend on the adiabaticity of the RSF-H resonance.
 
In the case of $Y_e = 0.51$, the dependence of the effective squared masses 
is different from other cases for the RSF-X and the RSF-L resonances.
At the high density limit, $\nu_e$ and $\bar{\nu}_e$ correspond to mass
eigenstates $\bar{\nu}_2$ and $\nu_3$, respectively.
Flavor eigenstates $\nu_{\mu,\tau}$ and $\bar{\nu}_{\mu,\tau}$ are the
mixed states of $\bar{\nu}_1$ and $\bar{\nu}_3$, and $\nu_1$ and $\nu_2$,
respectively.
The conversion between $\bar{\nu}_e$ and $\nu_{\mu,\tau}$ occurs 
at the RSF-H resonance.
RSF-L resonance appears at $\rho \sim 600$ ${\rm g cm^{-3}}$ and 
corresponds to the conversion of $\nu_e$ and $\bar{\nu}_{\mu,\tau}$.
The RSF-X resonance is found at $\rho \sim 1,300$ ${\rm g cm^{-3}}$.
This resonance leads to the conversion of $\nu_1$ and $\bar{\nu}_1$.
Since the corresponding mass eigenstates are different between the RSF-L 
and the RSF-X, these two resonances appear separately.
The densities and corresponding squared masses strongly depend on 
mass hierarchy as well as $Y_e$.

In order to find flavor transition at the RSF-H resonance, it is useful to
consider the two-flavor approximation.
First, we consider the transition between $\bar{\nu}_e$ and $\nu_{\tau}$.
The flavor change by the RSF-H is solved using the following equation:
\begin{equation}
i \frac{d}{dr}
\left( \begin{array}{c}
\bar{\nu}_e \\ \nu_{\tau}
\end{array} \right)
= \left( \begin{array}{cc}
\frac{\Delta m^2_{31}}{2 E_\nu}s_{13} -V_e & \mu_{e\tau}B_\bot \\ 
\mu_{e\tau}B_\bot & \frac{\Delta m^2_{31}}{2 E_\nu}c_{13} +V_\tau 
\end{array} \right)
\left( \begin{array}{c}
\bar{\nu}_e \\ \nu_{\tau}
\end{array} \right) .
\end{equation}
After calculating the eigenvalues of this equation, we derive the resonance
density of the transition between $\bar{\nu}_e$ and $\nu_{\tau}$ as
\begin{equation}
\rho_{res}(\bar{\nu}_e \leftrightarrow \nu_{\tau}) =
\frac{m_u \Delta m^2_{31} \cos2\theta_{13}}
{2 \sqrt{2} G_F E_\nu}
\frac{1}{1-2Y_e} .
\end{equation}
The resonance of the transition between $\bar{\nu}_e$ and $\nu_\tau$ exists
when the resonance density is larger than zero.
This condition holds in the case of normal mass hierarchy and $Y_e < 0.5$
[see Fig. 2(c) and Fig. 4] or the case of inverted mass hierarchy 
and $Y_e > 0.5$ [see Fig. 3(f) and Fig. 7].

\begin{figure*}
\includegraphics[angle=-90,width=6.7cm]{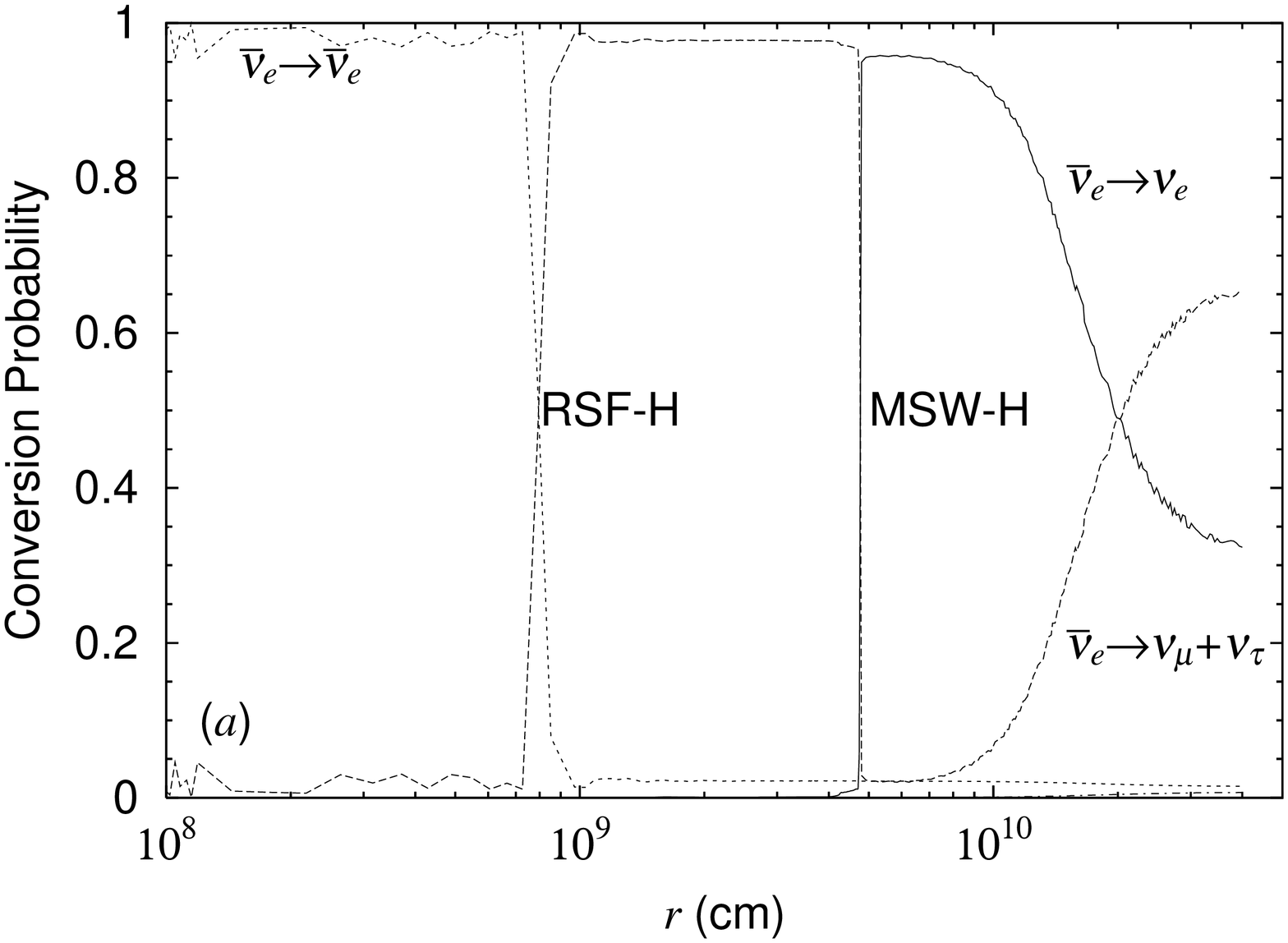}
\includegraphics[angle=-90,width=6.7cm]{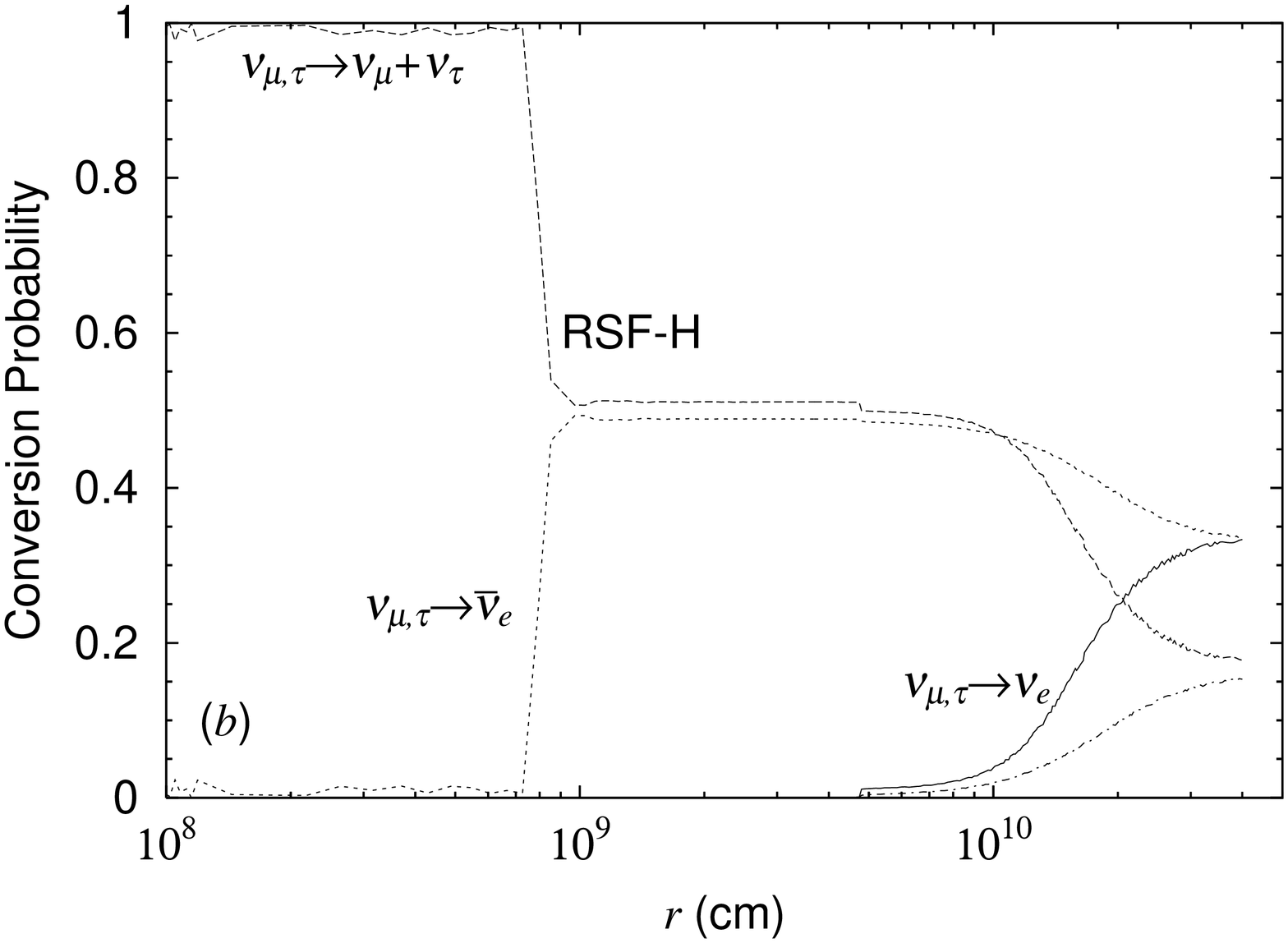}
\caption{\label{fig:prob_n49}
Conversion probabilities from $\bar{\nu}_e$ (a) and $\nu_{\mu,\tau}$ (b).
Mass hierarchy is normal.
Electron fraction of the inner region is set to be 0.49.
Solid lines, dotted lines, dashed lines, and dash-dotted lines correspond 
to the conversion probabilities to 
$\nu_e$, $\bar{\nu}_e$, $\nu_{\mu}+\nu_{\tau}$,
and $\bar{\nu}_{\mu}+\bar{\nu}_{\tau}$.
}
\end{figure*}
\begin{figure*}
\includegraphics[angle=-90,width=6.7cm]{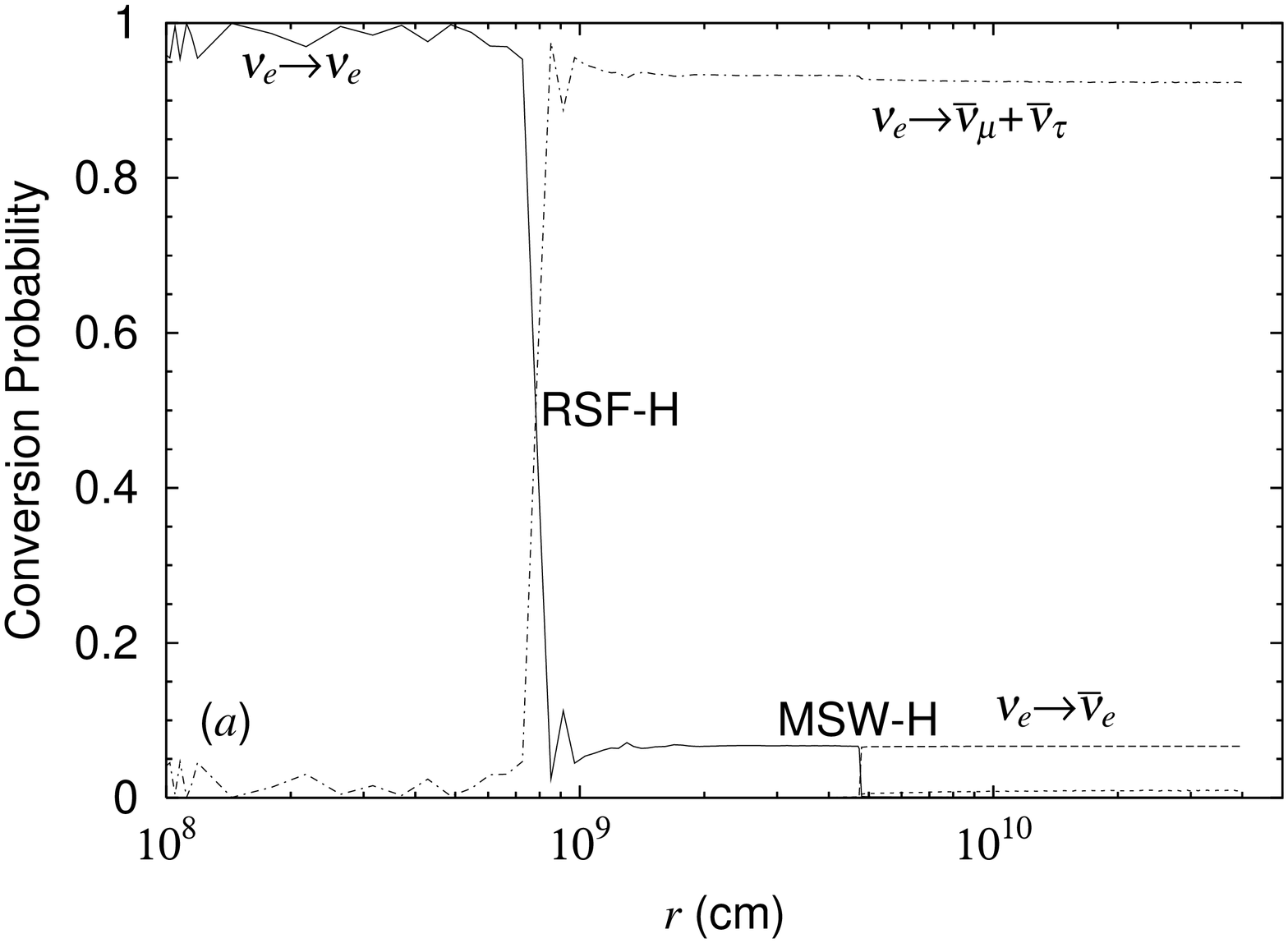}
\includegraphics[angle=-90,width=6.7cm]{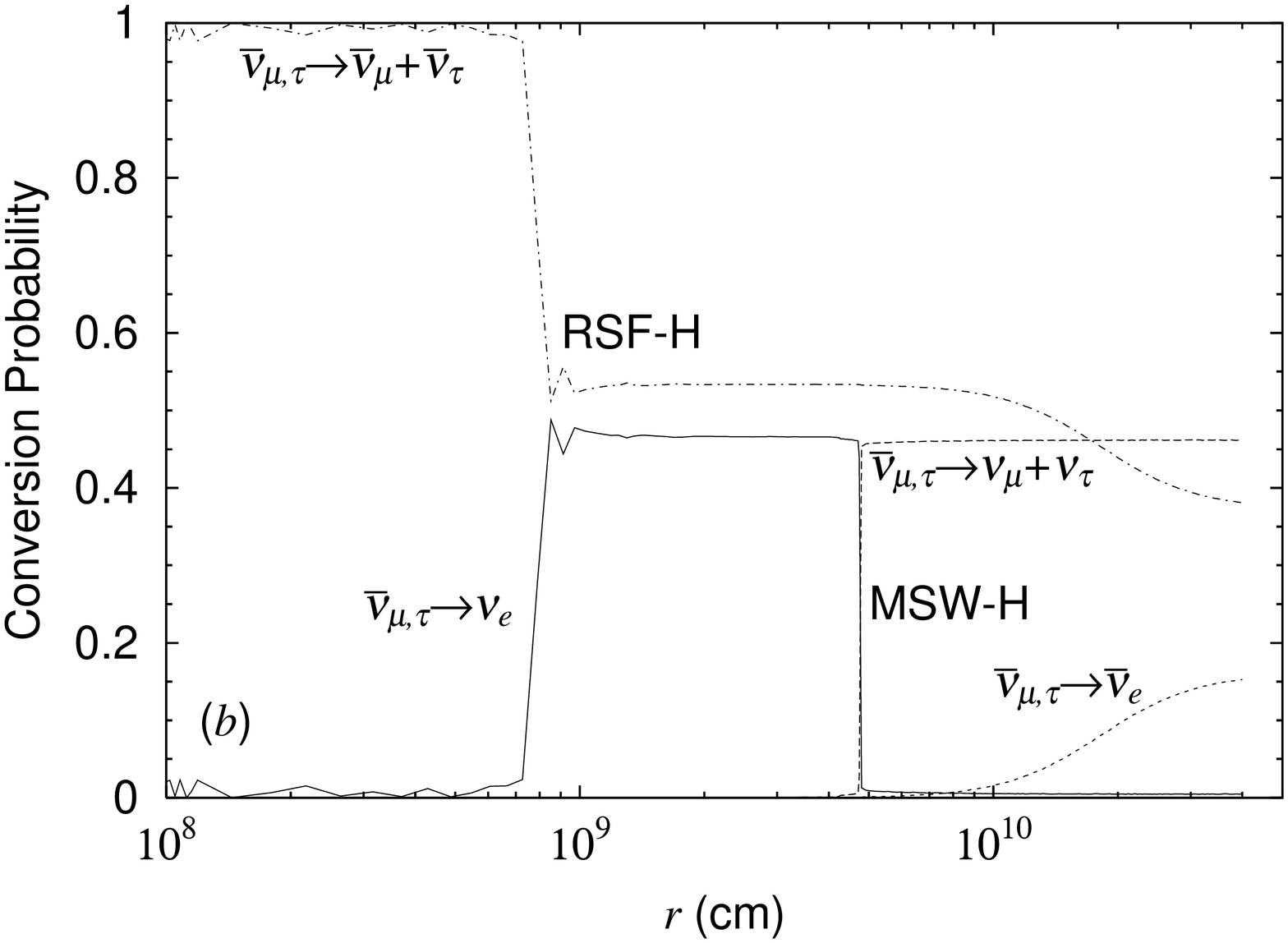}
\caption{\label{fig:prob_n51}
Conversion probabilities from $\nu_e$ (a) and $\bar{\nu}_{\mu,\tau}$ (b).
Mass hierarchy is normal.
Electron fraction of the inner region is set to be 0.51.
Solid lines, dotted lines, dashed lines, and dash-dotted lines correspond to 
the conversion probabilities to 
$\nu_e$, $\bar{\nu}_e$, $\nu_{\mu}+\nu_{\tau}$,
and $\bar{\nu}_{\mu}+\bar{\nu}_{\tau}$.
}
\end{figure*}

We can also discuss the condition of the transition between $\nu_e$ and
$\bar{\nu}_\tau$ by changing the sign of $V_e$ and $V_\tau$ in Eq. (20).
The obtained resonance density is
\begin{equation}
\rho_{res}(\nu_e \leftrightarrow \bar{\nu}_{\tau}) =
\frac{m_u \Delta m^2_{31} \cos2\theta_{13}}{2 \sqrt{2} G_F E_\nu}
\frac{1}{2Y_e-1} .
\end{equation}
This equation shows that the resonance of the transition between
$\nu_e$ and $\bar{\nu}_\tau$ exists in the case of normal mass hierarchy
and $Y_e > 0.5$ [see Fig. 2(f) and Fig. 5] or in the case of inverted mass
hierarchy and $Y_e < 0.5$ [see Fig. 3(c) and Fig. 6].

\subsection{Conversion probabilities}

We show the conversion probabilities of SN neutrinos passing
through the SN ejecta at 4 s after the core bounce.
We set the neutrino energy of 20 MeV, the mixing angle 
$\sin^22\theta_{13} = 0.04$ corresponding to adiabatic MSW-H resonance,
and $B_0 = 1 \times 10^{11}$ G.
The shock effect at the RSF-H resonance is quite small at that time 
and with the neutrino energy.
We assume that $Y_e = 0.49$ or 0.51 inside the mass coordinate of
$M_r = 1.43 M_\odot$, of which the radius is $r = 1.6 \times 10^9$ cm
at that time.

First, Fig. 4 shows the conversion probabilities from $\bar{\nu}_e$ [4(a)]
and $\nu_{\mu,\tau}$ [4(b)].
The obtained result is similar to that in \cite{as03a,as03b,as03c,af03}.
Almost all $\bar{\nu}_e$ change to $\nu_{\mu,\tau}$ at the RSF-H resonance
and convert to $\nu_e$ at MSW-H resonance.
Finally, 30 \% and 70 \% of the $\bar{\nu}_e$ become $\nu_e$ and 
$\bar{\nu}_{\mu,\tau}$, respectively.
On the other hand, a half of $\nu_{\mu,\tau}$ changes to $\bar{\nu}_e$
at the RSF-H resonance.
In vacuum, 30 \% and 70 \% of the $\nu_{\mu,\tau}$ become
$\nu_{\mu,\tau}$ and $\nu_e$, and 30 \% and 70 \% of the $\bar{\nu}_e$ 
become $\bar{\nu}_{\mu,\tau}$ and $\bar{\nu}_e$.
The conversion probabilities from $\nu_e$ and $\bar{\nu}_{\mu,\tau}$
are the same as in the case that the RSF conversions are not considered.

When we consider mass eigenstates, this conversion is explained more easily.
At high density limit, $\bar{\nu}_e$ corresponds to $\nu_2$.
The $\nu_2$ relates to the RSF-H, MSW-H, and MSW-L and these resonances 
are adiabatic.
Therefore, all $\bar{\nu}_e$ in high density are $\nu_2$ independent of
the density.
The $\nu_2$ corresponds to the mixed states of  70 \% of $\nu_{\mu,\tau}$
and 30 \% of $\nu_e$ in vacuum.
The conversion of $\nu_{\mu,\tau}$ is explained similarly.
The $\nu_{\mu,\tau}$ in high density correspond to the mixed states
of $\nu_1$ and $\bar{\nu}_1$.
The $\nu_1$ and $\bar{\nu}_1$ convert each other at RSF-L resonance.
However, we do not see any effects at the resonance.
MSW-L resonance may convert $\nu_1$ and $\nu_2$.
However, this resonance is adiabatic in this case.
In vacuum, $\nu_1$ is the mixed state of 70 \% of $\nu_e$ and 
30 \% of $\nu_{\mu,\tau}$.
The mass eigenstate $\bar{\nu}_1$ is the mixed state of 
70 \% of $\bar{\nu}_e$  and 30 \% of $\bar{\nu}_{\mu,\tau}$.

\begin{figure*}
\includegraphics[angle=-90,width=6.7cm]{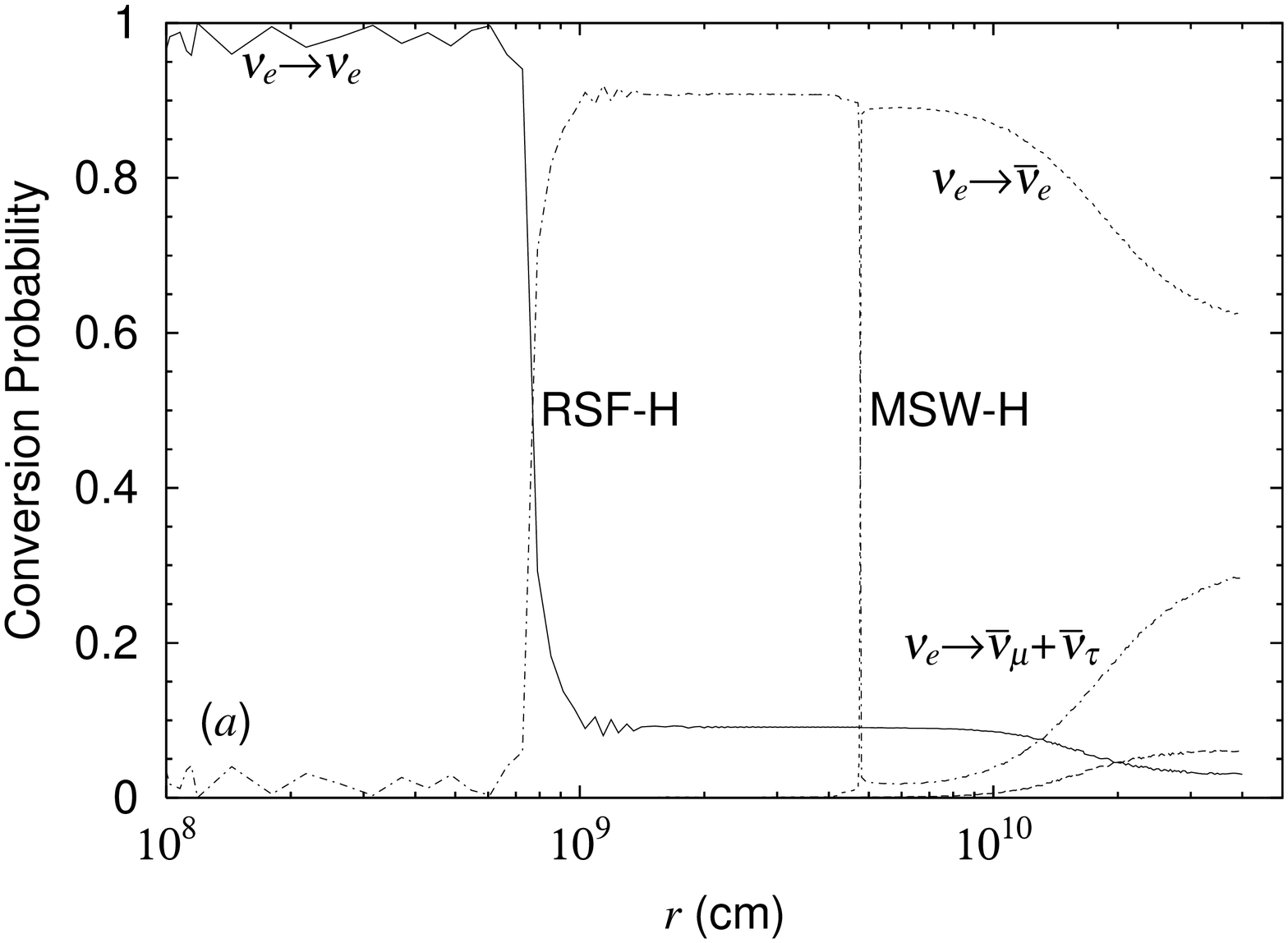}
\includegraphics[angle=-90,width=6.7cm]{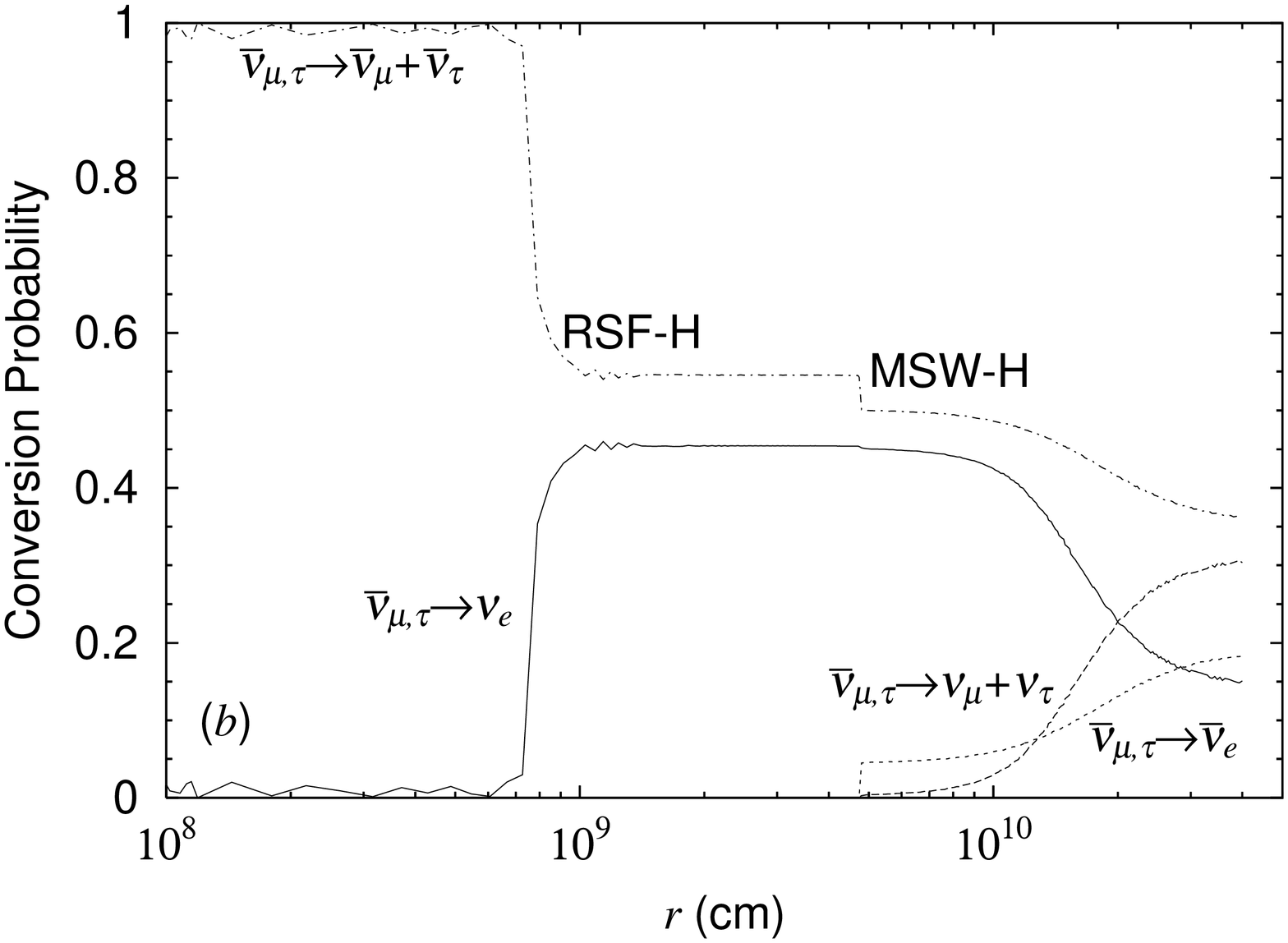}
\caption{\label{fig:prob_i49}
Conversion probabilities from $\nu_e$ (a) and $\bar{\nu}_{\mu,\tau}$ (b).
Mass hierarchy is inverted.
Electron fraction of the inner region is set to be 0.49.
Solid lines, dotted lines, dashed lines, and dash-dotted lines correspond to 
the conversion probabilities to 
$\nu_e$, $\bar{\nu}_e$, $\nu_{\mu}+\nu_{\tau}$,
and $\bar{\nu}_{\mu}+\bar{\nu}_{\tau}$.
}
\end{figure*}

\begin{figure*}
\includegraphics[angle=-90,width=6.7cm]{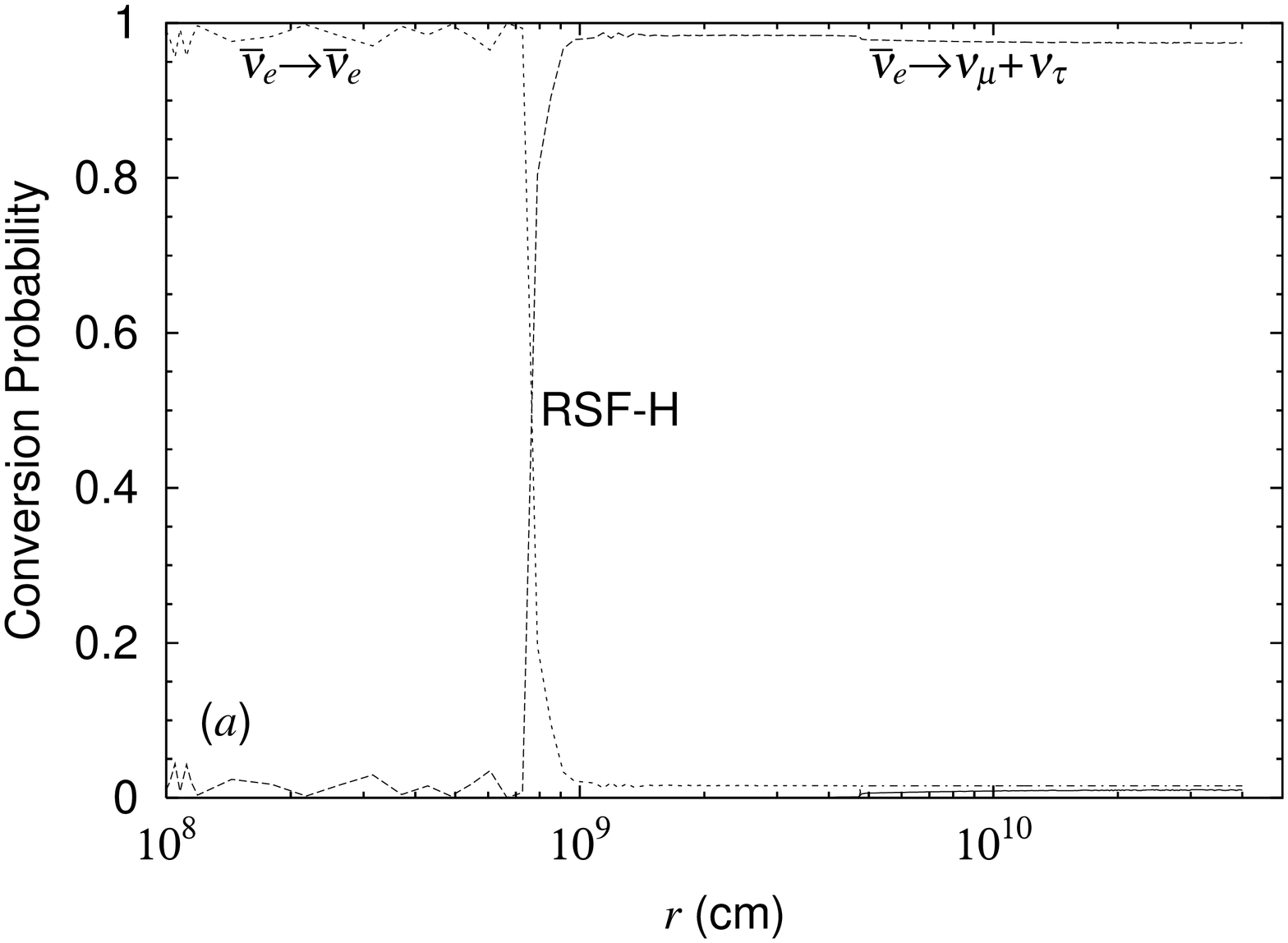}
\includegraphics[angle=-90,width=6.7cm]{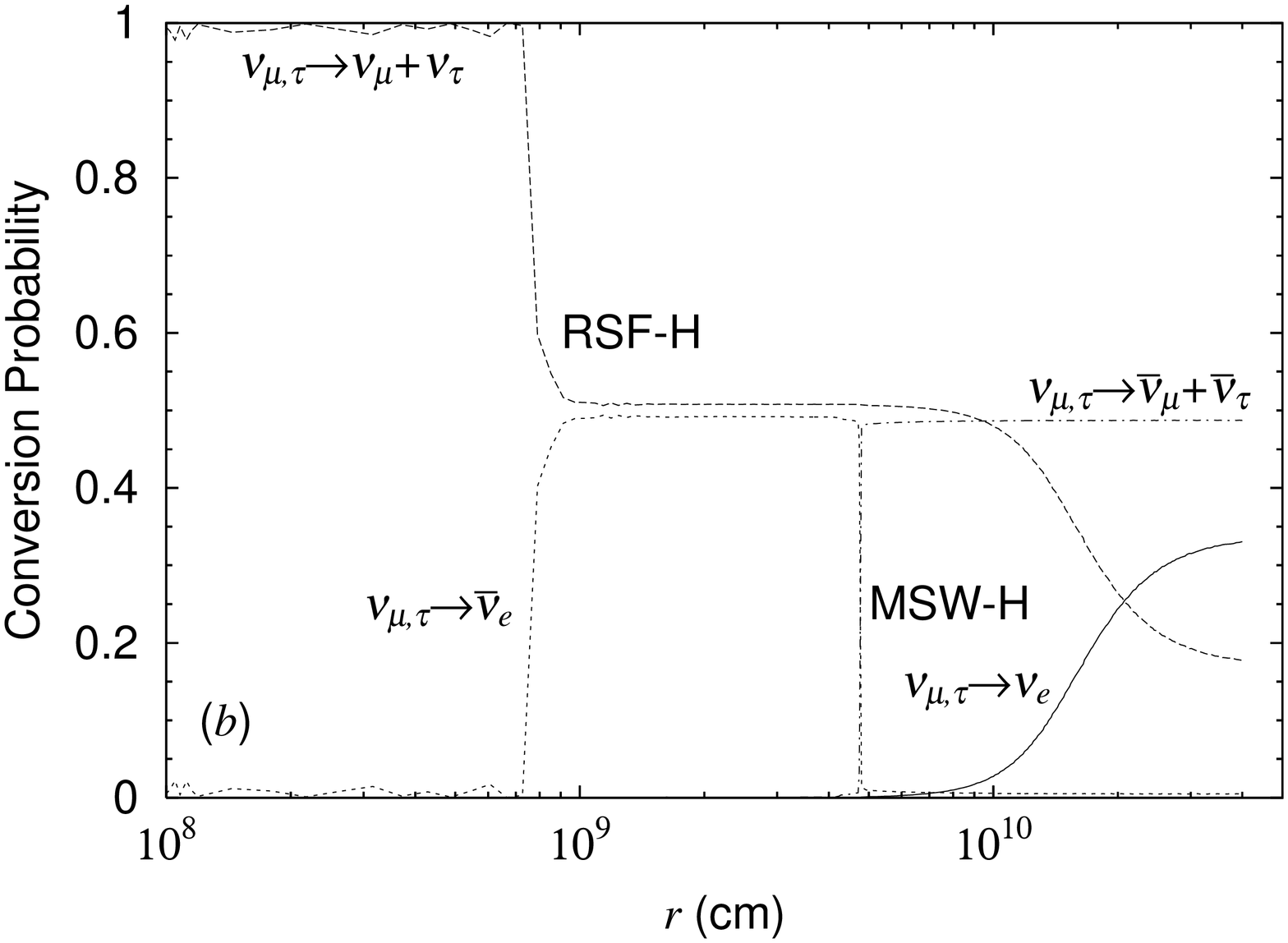}
\caption{\label{fig:prob_i51}
Conversion probabilities from $\bar{\nu}_e$ (a) and $\nu_{\mu,\tau}$ (b).
Mass hierarchy is inverted.
Electron fraction of the inner region is set to be 0.51.
Solid lines, dotted lines, dashed lines, and dash-dotted lines correspond to 
the conversion probabilities to 
$\nu_e$, $\bar{\nu}_e$, $\nu_{\mu}+\nu_{\tau}$,
and $\bar{\nu}_{\mu}+\bar{\nu}_{\tau}$.
}
\end{figure*}

Second, Fig. 5 shows the conversion probabilities from $\nu_e$ [5(a)] and 
$\bar{\nu}_{\mu,\tau}$ [5(b)] in the case of the normal mass hierarchy and 
$Y_e = 0.51$ in the inner region.
We see that the conversion between $\nu_e$ and $\bar{\nu}_{\mu,\tau}$
occurs at RSF-H resonance.
The conversions at MSW-H and MSW-L resonances are also shown.
However, no flavor conversion occurs at RSF-L resonance.
In this case, all $\nu_e$ corresponds to $\bar{\nu}_3$ at the high density
limit.
When $Y_e$ changes from $Y_e > 0.5$ to $Y_e < 0.5$, no conversion of
mass eigenstates occurs for $\bar{\nu}_3$.
In the density region lower than MSW-H resonance, $\bar{\nu}_3$ corresponds
to $\bar{\nu}_{\mu,\tau}$ and a small fraction of $\bar{\nu}_e$.
Thus, $\nu_e$ changes to $\bar{\nu}_{\mu,\tau}$.
On the other hand, $\bar{\nu}_{\mu,\tau}$ are the mixed states of
$\bar{\nu}_2$ and $\nu_3$.
The conversion between $\bar{\nu}_2$ and $\nu_3$ occurs at the location where 
the $Y_e$ changes.
Then the two eigenstates convert again at nonadiabatic RSF-X
resonance.
Thus, we see no effects in the conversions between $\bar{\nu}_2$
and $\nu_3$.
The $\nu_3$ corresponds to $\nu_{\mu,\tau}$ and a small fraction of $\nu_e$.
The $\nu_2$ is the mixed flavor eigenstates of  70 \% of $\bar{\nu}_{\mu,\tau}$
and 30 \% of $\bar{\nu}_e$ in vacuum.

Third, we consider the case of the inverted mass hierarchy and $Y_e = 0.49$
in the inner region.
Figure 6 shows the conversion probabilities from $\nu_e$ [6(a)] and 
$\bar{\nu}_{\mu,\tau}$ [6(b)].
RSF-H conversion occurs between $\nu_e$ and $\bar{\nu}_{\mu,\tau}$ similarly
to the second case.
Flavor conversion occurs at RSF-H and MSW-H resonances for the original 
$\nu_e$.
It occurs at RSF-H and MSW-L resonances for the original 
$\bar{\nu}_{\mu,\tau}$.

The flavor eigenstate $\nu_e$ corresponds to $\nu_1$ at high density limit.
The mass eigenstate $\nu_1$ does not convert at RSF-H resonance.
It converts to $\bar{\nu}_1$ at RSF-L resonance because of nonadiabaticity
of this resonance.
Therefore, the neutrinos originally produced as $\nu_e$ appear as 
$\bar{\nu}_1$, corresponding to the mixtures of 70 \% of $\bar{\nu}_e$ and 
30 \% of $\bar{\nu}_{\mu,\tau}$ in vacuum.
The flavor eigenstate $\bar{\nu}_{\mu,\tau}$ is a mixed state of 
$\nu_2$ and $\bar{\nu}_2$.
There are no resonances for $\bar{\nu}_2$.
The $\nu_2$ has RSF-H and MSW-L resonances.
Since both of the resonances are adiabatic, the mixed state of $\nu_2$
and $\bar{\nu}_2$ does not change the mass eigenstate throughout the
spectra.
The $\nu_2$ ($\bar{\nu}_2$) corresponds to 70 \% of $\nu_{\mu,\tau}$ 
($\bar{\nu}_{\mu,\tau}$) in vacuum.


Finally, we show the case of the inverted mass hierarchy and $Y_e = 0.51$
in the inner region.
Figure 7 shows the conversion probabilities from $\bar{\nu}_e$ [7(a)] and 
$\nu_{\mu,\tau}$ [7(b)].
The conversion between $\bar{\nu}_e$ and $\nu_{\mu,\tau}$ occurs at RSF-H
resonance in this case.
This is similar to the first case.
Almost all $\bar{\nu}_e$ convert to $\nu_{\mu,\tau}$ at the RSF-H.
The $\bar{\nu}_e$ corresponds to $\nu_3$ at the high density limit and
there is no resonance after the RSF-H resonance.
On the other hand, $\nu_{\mu,\tau}$ changes flavors three times.
The $\nu_{\mu,\tau}$ is a mixed state of $\bar{\nu}_1$ and $\bar{\nu}_3$
at high density limit.
About a half of $\nu_{\mu,\tau}$, corresponding to $\bar{\nu}_3$, converts
to $\bar{\nu}_e$ at RSF-H resonance.
The mass eigenstate $\bar{\nu}_3$ changes the flavors at MSW-H resonance.
It appears as the mixed state of $\bar{\nu}_{\mu,\tau}$ and a small fraction
of $\bar{\nu}_e$ in vacuum.
The mass eigenstate $\bar{\nu}_1$ converts to $\nu_1$ at RSF-X resonance, 
so that it appears as 70 \% of $\nu_e$ and 30 \% of $\nu_{\mu,\tau}$ 
in vacuum.

\subsection{Neutrino signal}

\begin{figure*}
\includegraphics[angle=-90,width=6.7cm]{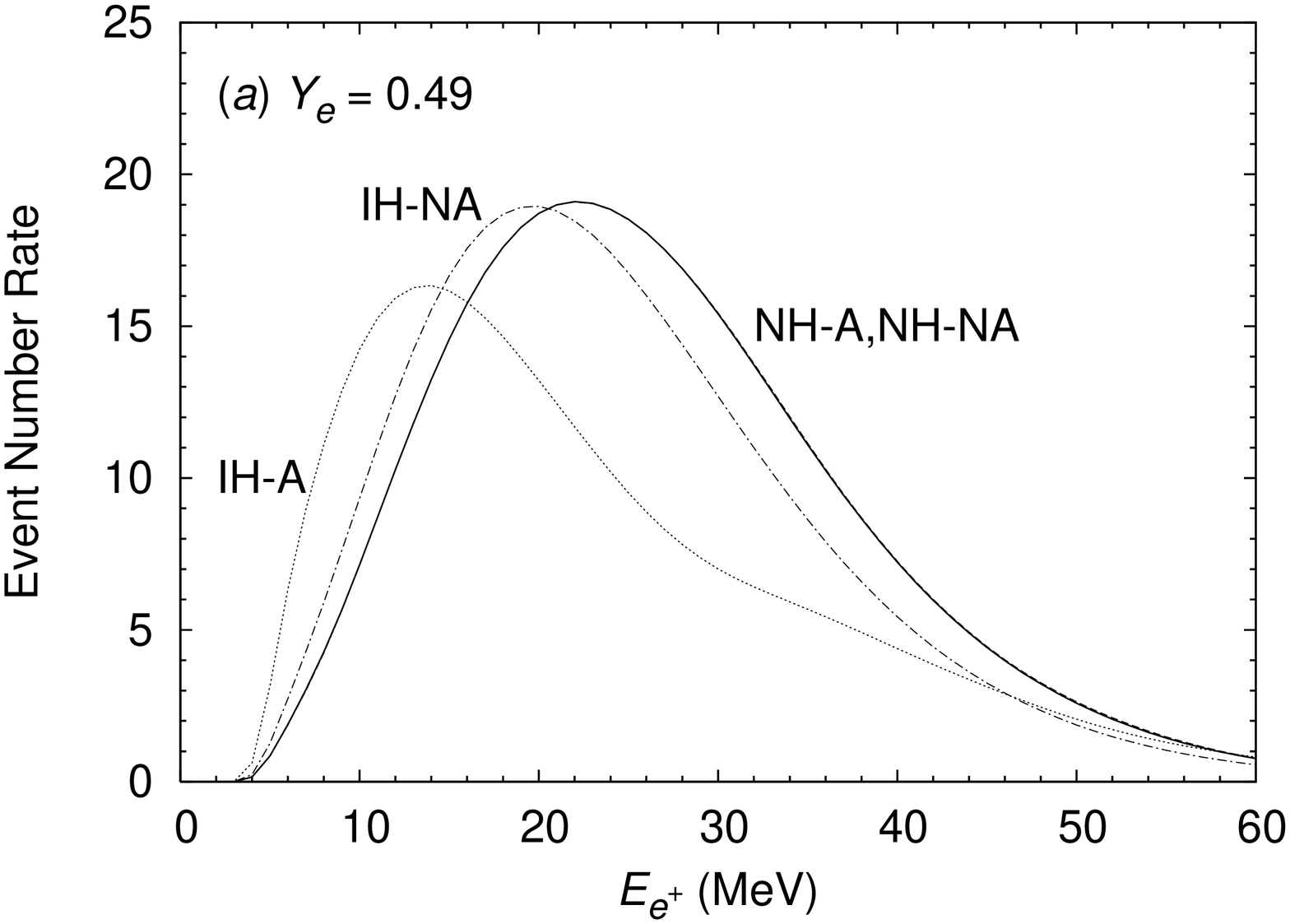}
\includegraphics[angle=-90,width=6.7cm]{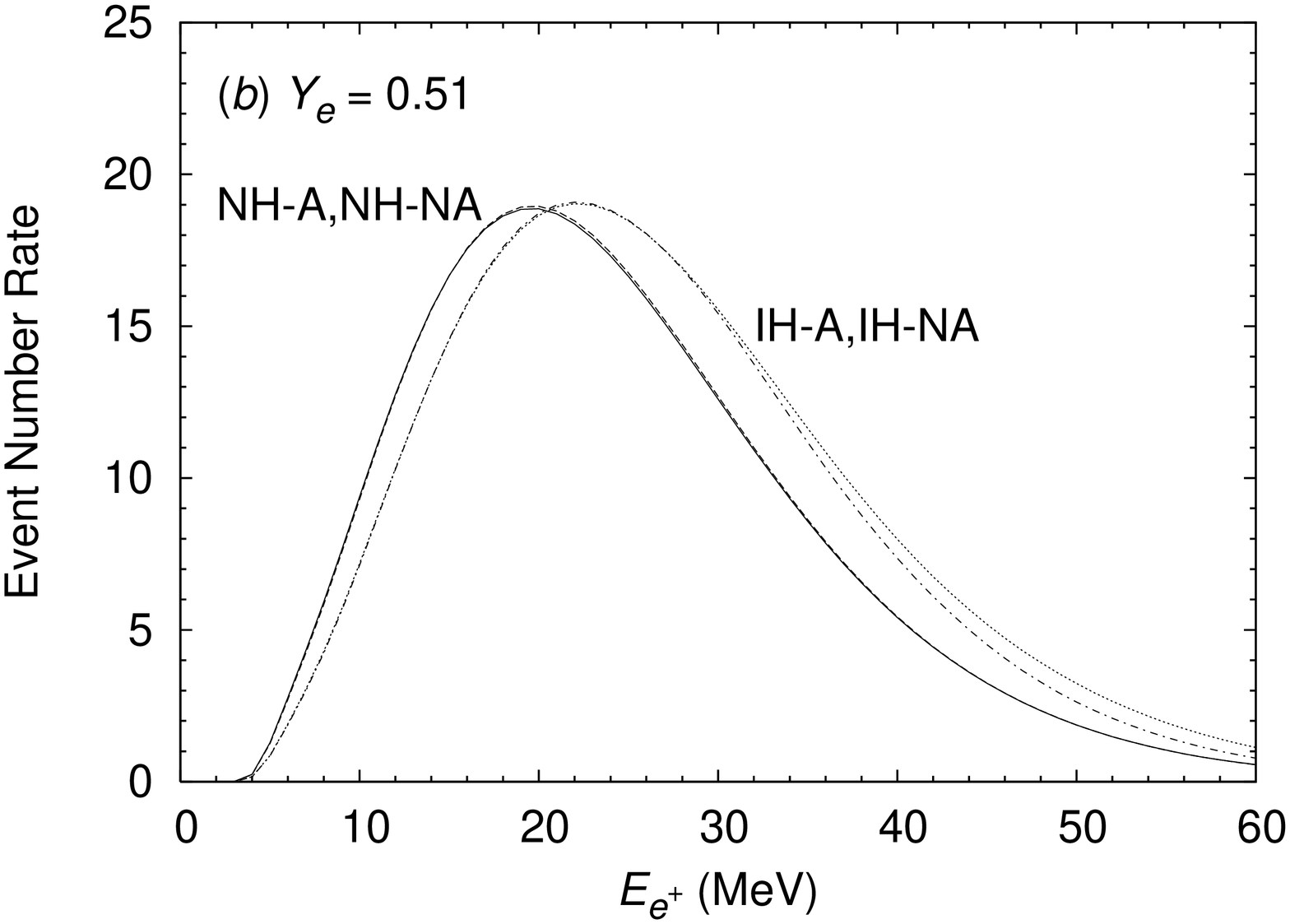}
\includegraphics[angle=-90,width=6.7cm]{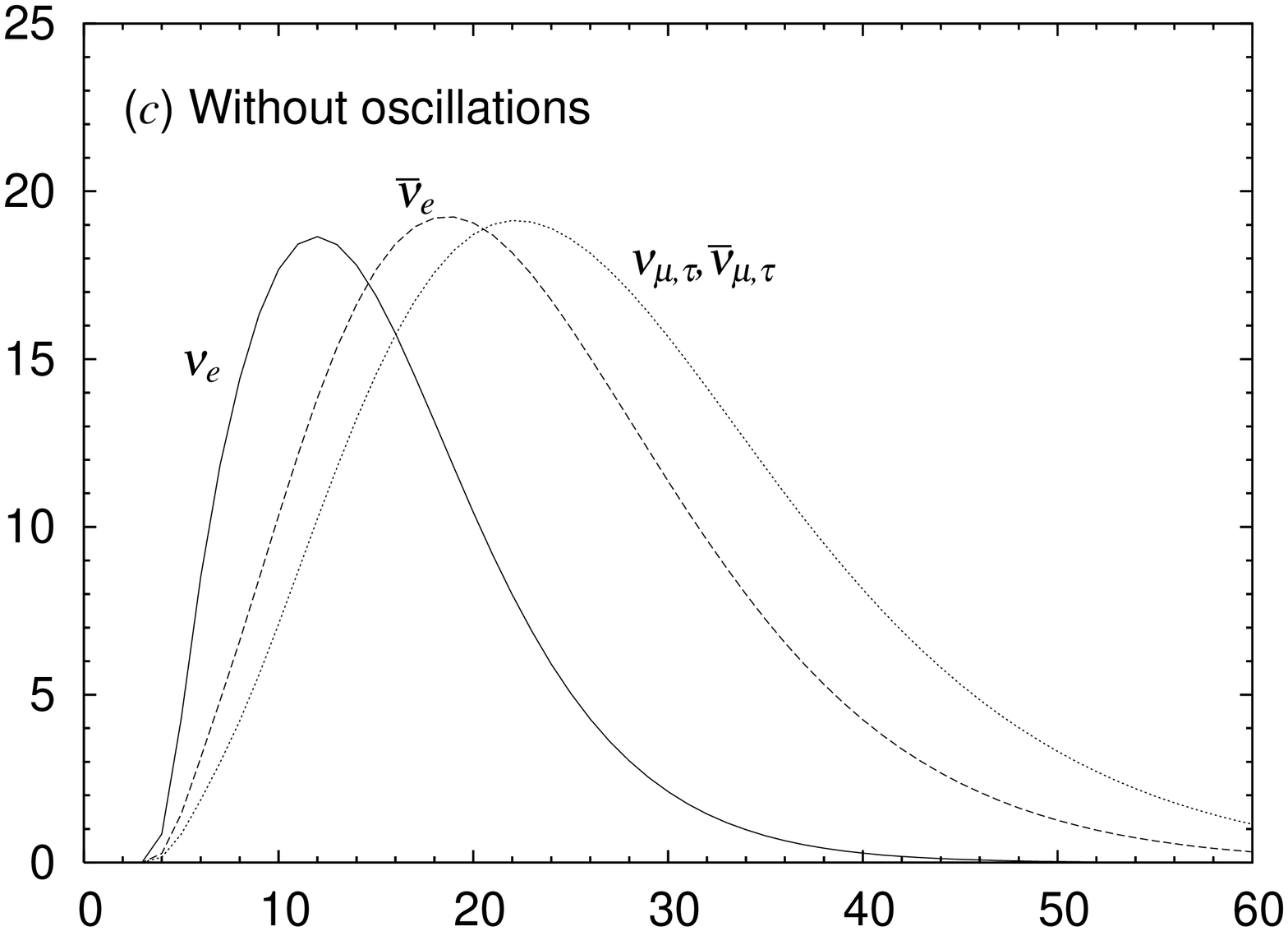}
\caption{\label{fig:spectra_nuebp}
The positron event number rate by $p(\bar{\nu}_e,e^+)n$ as a function of 
emitted positron energy at $t = 4$ s in the cases of $Y_e = 0.49$
(a) and 0.51 (b) at the inner region of the SN ejecta.
Solid lines, dashed lines, dotted lines, and dash-dotted lines are 
the cases of NH-A, NH-NA, IH-A, and IH-NA, respectively.
(c): The positron event number rates corresponding to Fermi distributions
with the temperature of 3.2 MeV (solid line), 5.0 MeV (dashed line), and
6.0 MeV (dotted line).
}
\end{figure*}

\begin{table*}
\caption{\label{tab:table1}
The final energy spectra of $\bar{\nu}_e$ [$\phi_{\bar{\nu}_e}$(fin)] 
with the relation of the original neutrino flux $\phi_{\nu_\alpha}$.
The event ratios $r_{{\rm L/H}}$ are classified into three patterns: 
(A) large ratio, (B) intermediate ratio, and (C) small ratio.
}
\begin{ruledtabular}
\begin{tabular}{lcc}
 & $\phi_{\bar{\nu}_e}$(fin) & Pattern \\
\hline
\multicolumn{3}{c}{MSW effect} \\
\hline
NH-A &
$|U_{e1}|^2 \phi_{\bar{\nu}_e} + (1-|U_{e1}|^2) \phi_{\bar{\nu}_x}$ & (B) \\
IH-A &
$|U_{e3}|^2 \phi_{\bar{\nu}_e} + (1-|U_{e3}|^2) \phi_{\bar{\nu}_x}$ & (C) \\
NA-NA and IA-NA &
$|U_{e1}|^2 \phi_{\bar{\nu}_e} + (1-|U_{e1}|^2) \phi_{\bar{\nu}_x}$ & (B) \\
\hline
\multicolumn{3}{c}{RSF conversion ($Y_e < 0.5$) and MSW effect} \\
\hline
NH-A &
$|U_{e1}|^2 \phi_{\nu_x} + |U_{e2}|^2 \phi_{\bar{\nu}_x}
+ |U_{e3}|^2 \phi_{\bar{\nu}_x}$ & (C) \\
NH-NA & 
$|U_{e1}|^2 \phi_{\nu_x} + |U_{e2}|^2 \phi_{\bar{\nu}_x}
+ |U_{e3}|^2 \phi_{\bar{\nu}_x}$ & (C) \\
IH-A & 
$|U_{e1}|^2 \phi_{\nu_e} + |U_{e2}|^2 \phi_{\bar{\nu}_x}
+ |U_{e3}|^2 \phi_{\bar{\nu}_e}$ & (A) \\
IH-NA & 
$|U_{e1}|^2 \phi_{\bar{\nu}_e} + |U_{e2}|^2 \phi_{\bar{\nu}_x}
+ |U_{e3}|^2 \phi_{\nu_e}$ & (B) \\
\hline
\multicolumn{3}{c}{RSF conversion ($Y_e > 0.5$) and MSW effect} \\
\hline
NH-A & 
$|U_{e1}|^2 \phi_{\bar{\nu}_e} + |U_{e2}|^2 \phi_{\bar{\nu}_x}
+ |U_{e3}|^2 \phi_{\nu_e}$ & (B) \\
NH-NA & 
$|U_{e1}|^2 \phi_{\bar{\nu}_e} + |U_{e2}|^2 \phi_{\bar{\nu}_x}
+ |U_{e3}|^2 \phi_{\nu_e}$ & (B) \\
IH-A & 
$|U_{e1}|^2 \phi_{\bar{\nu}_x} + |U_{e2}|^2 \phi_{\bar{\nu}_x}
+ |U_{e3}|^2 \phi_{\nu_x}$ & (C) \\
IH-NA & 
$|U_{e1}|^2 \phi_{\nu_x} + |U_{e2}|^2 \phi_{\bar{\nu}_x}
+ |U_{e3}|^2 \phi_{\bar{\nu}_x}$ & (C) \\
\end{tabular}
\end{ruledtabular}
\end{table*}

SN neutrinos change their flavors by the RSF conversion and the MSW effect
in SN ejecta.
The flavor changes depend on the distributions of the density and electron 
fraction.
Here we investigate the dependence of neutrino energy spectra and the
neutrino signals on the neutrino oscillation parameters and the electron
fraction.

\begin{figure*}
\includegraphics[width=6.7cm]{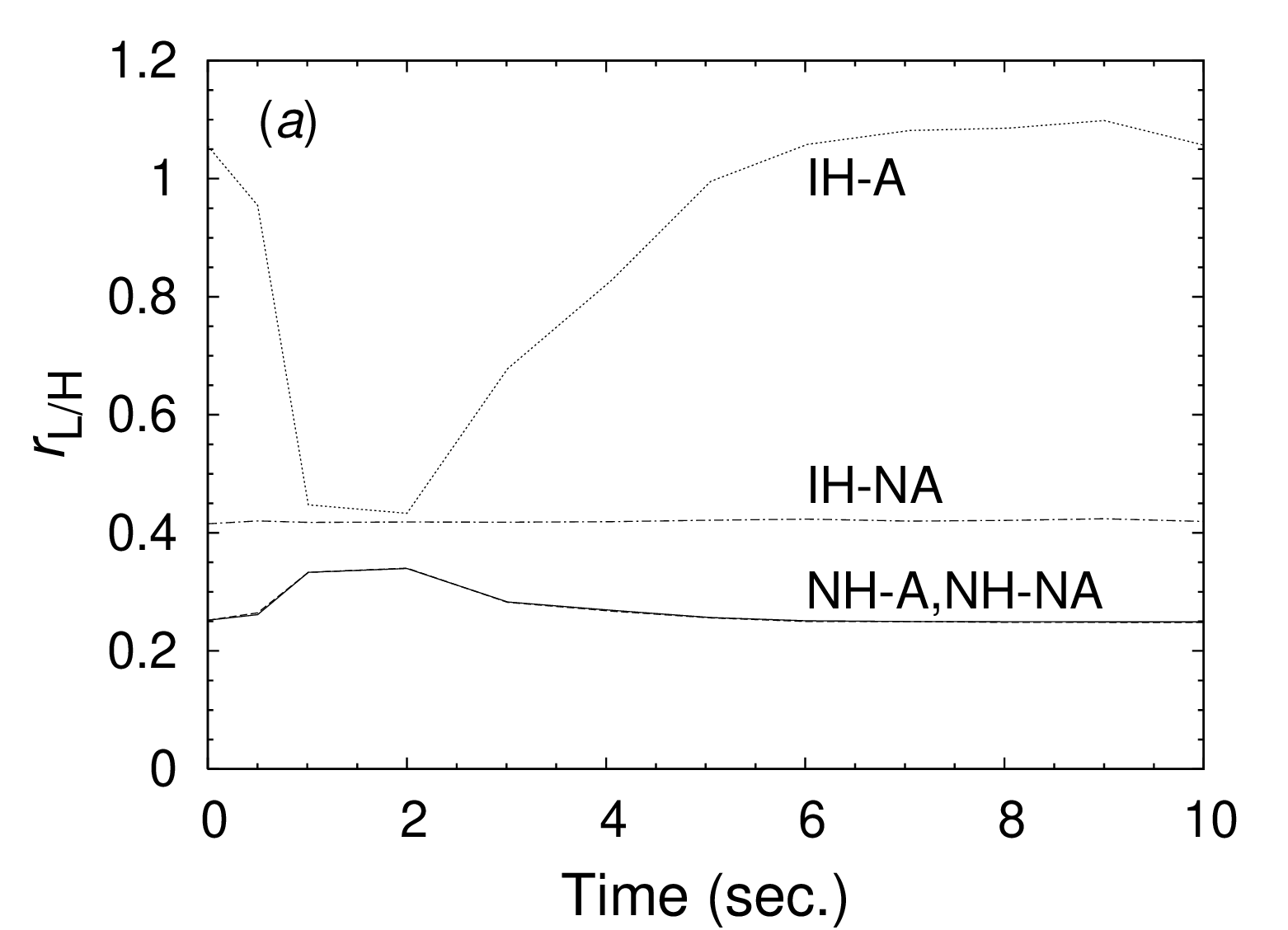}
\includegraphics[width=6.7cm]{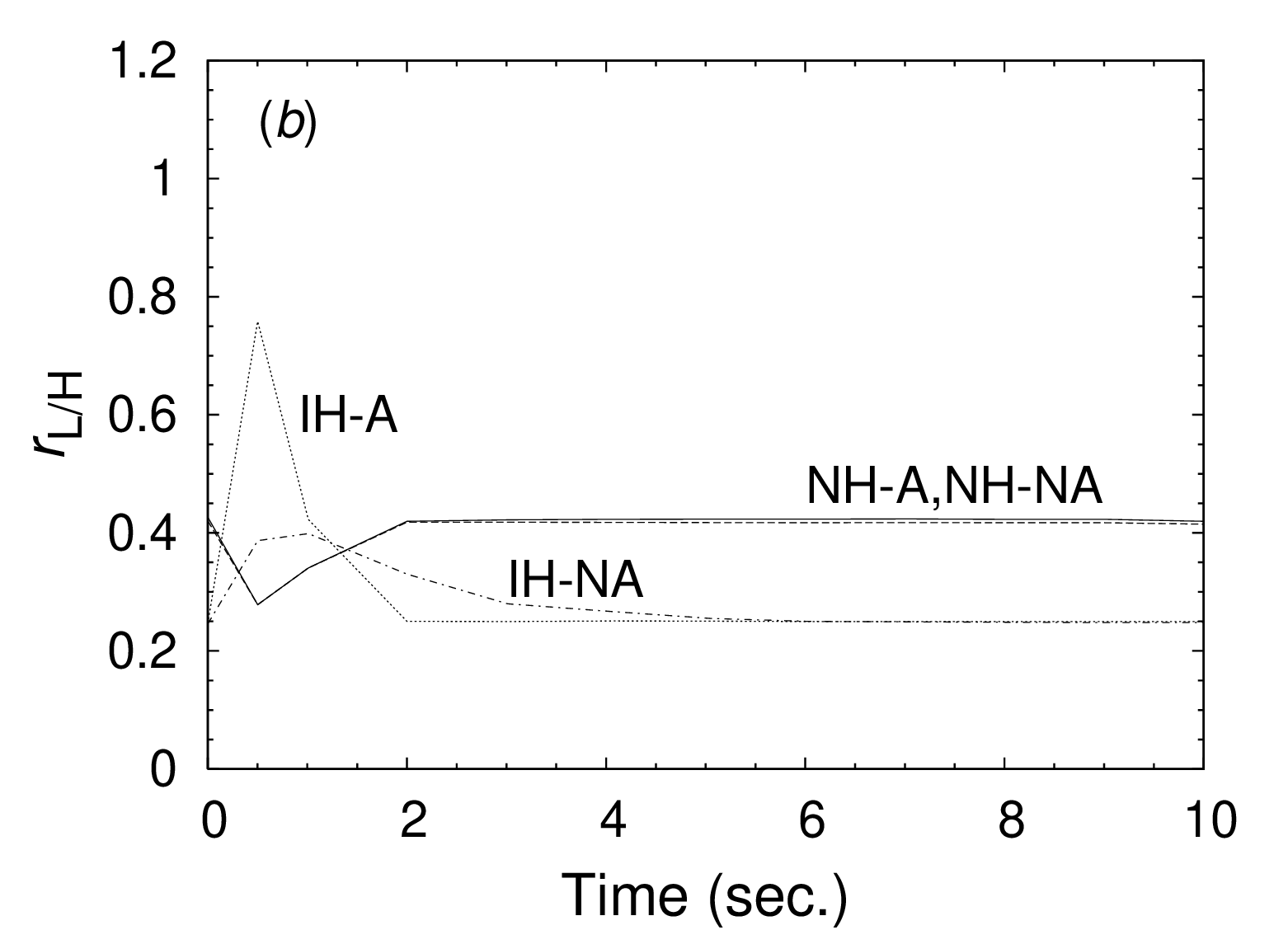}
\caption{\label{fig:rLH_tfix}
The time evolution of the event number ratio $r_{L/H}$ for
$p(\bar{\nu}_e,e^+)n$ with 22.5 kton water-{\v C}erenkov detector 
in the cases of $Y_e = 0.49$ ($a$) and $Y_e = 0.51$ ($b$) 
at the inner region of the SN ejecta.
RSF-H resonance is set to be adiabatic.
Solid lines, dashed lines, dotted lines, and dash-dotted lines are 
the cases of NH-A, NH-NA, IH-A, and IH-NA, respectively.
}
\end{figure*}

In this section, we consider the dependence on the mass hierarchies, electron
fraction in the inner region, and the adiabaticity of the MSW resonance.
We consider the two cases of $Y_e$ value equal to 0.49 and 0.51.
First, we show the energy spectra of the $\bar{\nu}_e$ signals at $t = 4$ s
with a 22.5 kton water {\v C}erenkov detector.
Figure 8 shows the positron energy spectra of the neutrino event rate by
$p(\bar{\nu}_e,e^+)n$ at $t = 4$ s.
In the case of $Y_e = 0.49$, the energy spectra are classified into
three patterns.
The lowest $\bar{\nu}_e$ energy spectrum is seen in the case of
the inverted mass hierarchy and adiabatic MSW-H resonance.
The intermediate energy spectrum is obtained in the inverted mass hierarchy
and nonadiabatic MSW-H resonance.
In the normal mass hierarchy, the $\bar{\nu}_e$ energy spectra indicate
the highest average energy independent of the adiabaticity of MSW-H resonance.
The energy spectra are almost identical to those of $\nu_{\mu,\tau}$
without neutrino oscillations.

On the other hand, in the case of $Y_e = 0.51$, the energy spectra
are classified into two patterns.
In the normal mass hierarchy, the energy spectra are almost identical
to the intermediate one in the case of $Y_e = 0.49$.
In the inverted mass hierarchy, they are almost identical to that of 
$\nu_{\mu,\tau}$ without neutrino oscillations.
We do not see the dependence on $\sin^22\theta_{13}$.

The final neutrino signal can be calculated using transition probabilities
of antineutrinos.
Table I shows the relation of the final neutrino spectra to the original
ones.
The final $\bar{\nu}_e$ spectrum $\phi_{\bar{\nu}_e}$(fin) does not depend
on the adiabaticity of MSW-H resonance except in the case of inverted mass 
hierarchy and $Y_e = 0.49$.
In the case of the inverted mass hierarchy, adiabatic MSW-H resonance,
and $Y_e = 0.51$, the flavor change between $\nu_{\mu,\tau}$ and 
$\bar{\nu}_{\mu,\tau}$ occurs but their energy spectra are identical.

We briefly discuss the event number rate of $p(\bar{\nu}_e,e^+)n$.
The event number rate decreases with time exponentially.
This reflects the assumption that the neutrino luminosity exponentially
decreases with time.
In $t \sim 0.5-2$ s the shock effect also exhibits in the event number rate.
The dependence of the event rate at a given time on neutrino oscillation
parameters and the electron fraction is similar to the dependence of the
energy spectra.

In the case of $Y_e = 0.49$, the event rates are classified into three
patterns.
The first pattern is the largest event rate in the normal mass hierarchy.
The second pattern is a smaller event rate in the inverted mass hierarchy 
and the nonadiabatic MSW-H resonance .
The third pattern is the smallest event rate in the inverted mass hierarchy 
and the adiabatic MSW-H resonance.
The event number rate is about $2200-3500$ events s$^{-1}$ at $t = 0$ s
and decreases to $70-120$ events s$^{-1}$ at $t = 10$ s.
When the shock wave passes through the region of RSF-H resonance, 
the event rates change.
The event rates decrease in the normal mass hierarchy.
On the other hand, in the inverted mass hierarchy and adiabatic RSF-H
resonance, the event rates rise.

In the case of $Y_e = 0.51$, the event rates are classified into two
patterns.
The event rates are small in the normal mass hierarchy and large in the
inverted mass hierarchy.
The dependence on the adiabaticity of the MSW-H resonance is not seen.
The difference in the event rates between the normal mass hierarchy and
the inverted mass hierarchy is smaller than the difference in $Y_e = 0.49$.
The event rates are about 3000 events s$^{-1}$ at $t = 0$ s
and about 100 events s$^{-1}$ at $t = 10$ s.
The shock effect is also seen in $t \sim 0.5-2$ s.
The event rates increase in the normal mass hierarchy and decrease
in the inverted mass hierarchy.

\begin{figure*}
\includegraphics[width=6.7cm]{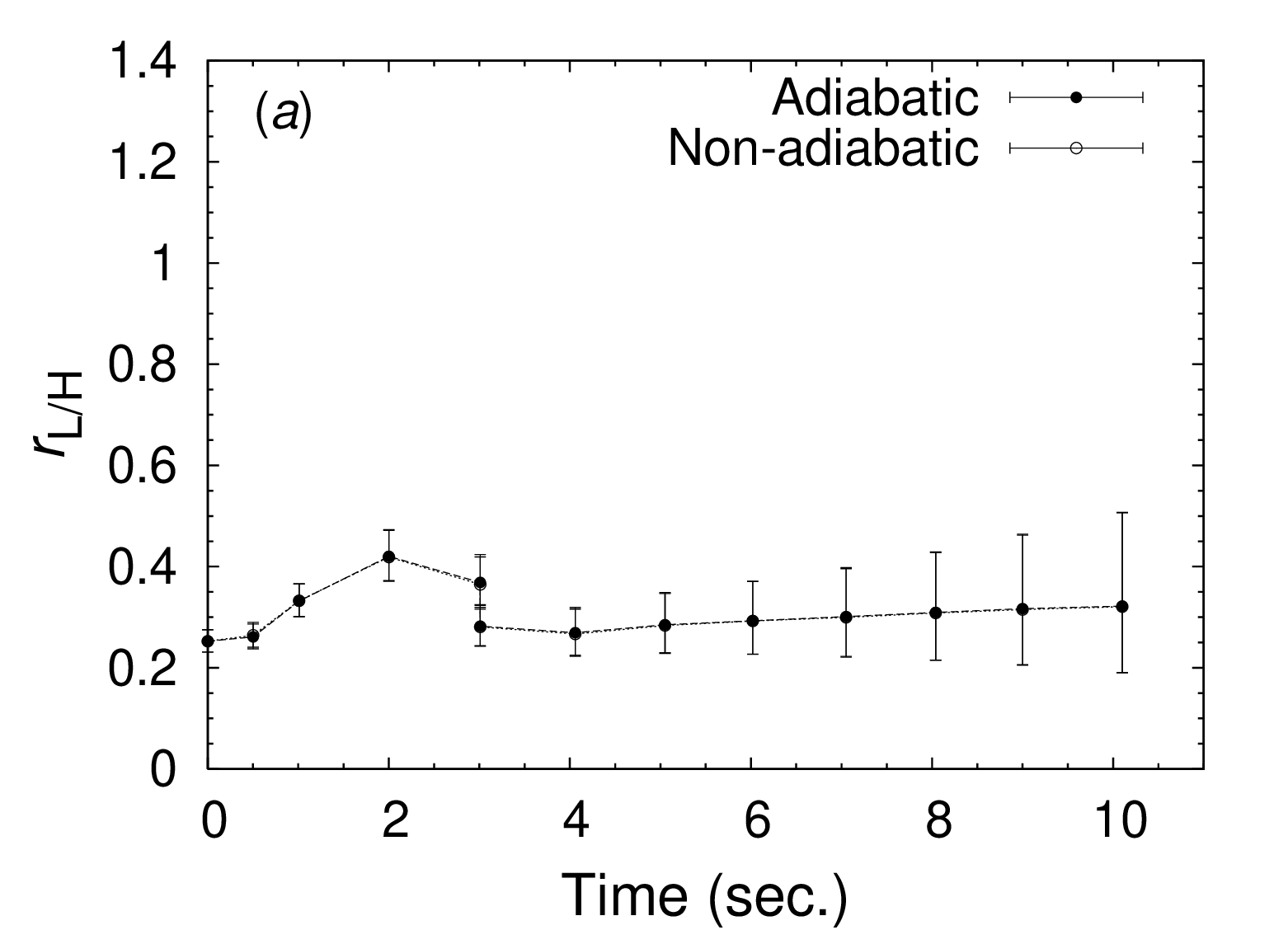}
\includegraphics[width=6.7cm]{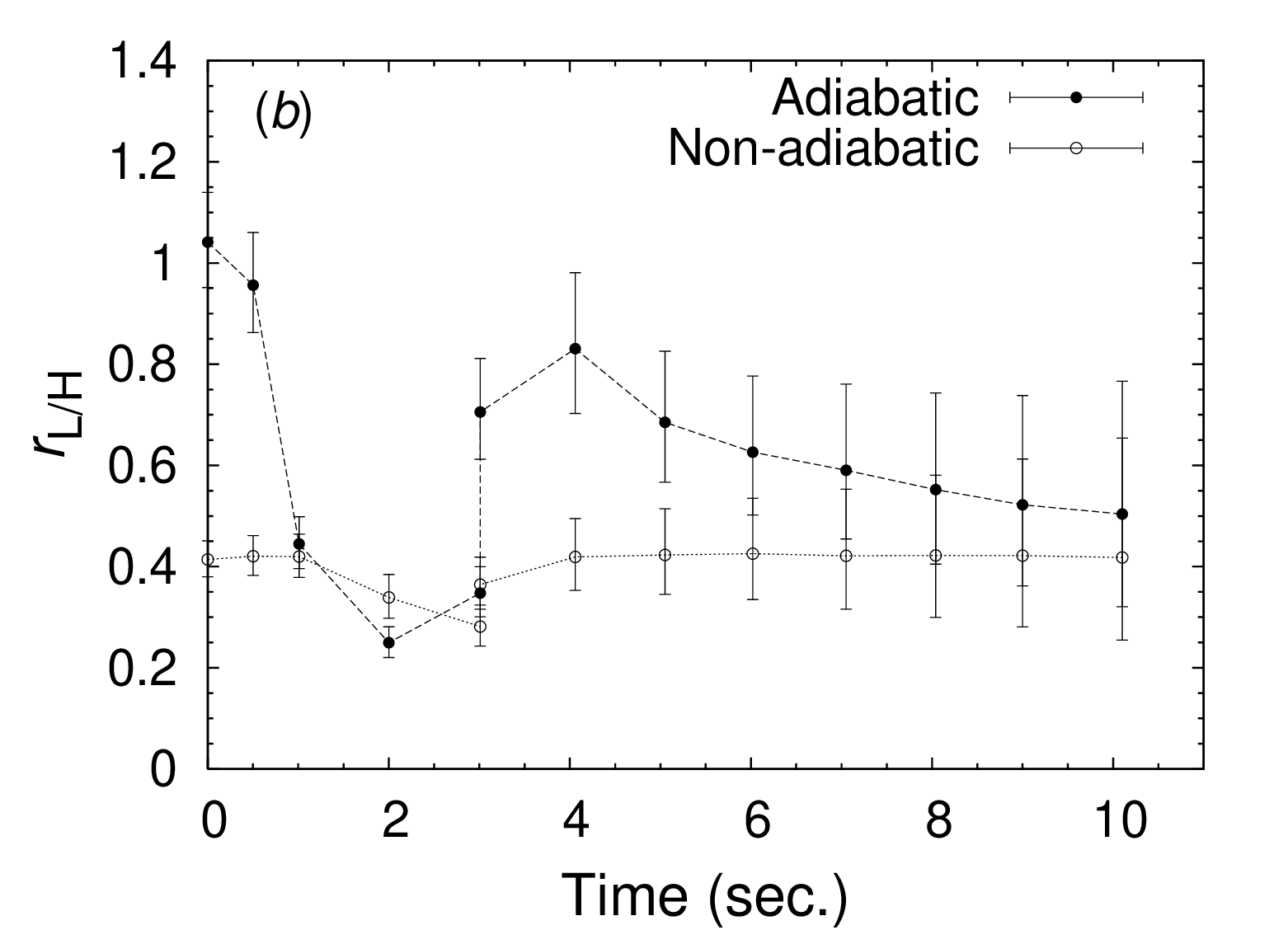}
\caption{\label{fig:rLH_tevol}
The time evolution of positron event number ratio $r_{{\rm L/H}}$ 
by $p(\bar{\nu}_e,e^+)n$ with 22.5 kton water-{\v C}erenkov detector 
for normal ($a$) and inverted ($b$) mass hierarchy.
Closed and open circles correspond to the cases of $\sin^22\theta_{13}$
= 0.04 and $1 \times 10^{-6}$, respectively.
Error bars are evaluated from the root of event numbers 
(see text for details).
}
\end{figure*}

The event ratio of the low energy component to the high energy component of
the neutrino signal is a useful measure to see the difference
by neutrino oscillation parameters.
We investigate the dependence of the $\bar{\nu}_e$ event ratio 
$r_{{\rm L/H}}$ on the oscillation parameters and the electron fraction 
in the water-{\v C}erenkov detector.
Here we define the event ratio $r_{{\rm L/H}}$ as
\begin{equation}
r_{{\rm L/H}} = 
\frac{\mbox{Event number for $E_e < 15$ MeV}}
{\mbox{Event number for $E_e > 25$ MeV}}.
\end{equation}
We also investigate the time dependence, i.e., the shock propagation effect.
However, we do not take account of the time evolution of $Y_e$ 
in the inner region.
We will discuss this effect by the $Y_e$ evolution in Sec. IV.

We consider the case of $Y_e = 0.49$ (see the left-hand panel of Fig. 9).
We see three patterns of the time evolution of the event ratios as shown
in Table 1.
First, the large ratio [pattern (A)], $r_{{\rm L/H}} \sim 1.0$, 
is seen in the inverted mass hierarchy and adiabatic MSW-H resonance.
In this case $\phi_{\bar{\nu}_e}$(fin) is a mixed spectrum of 
$\sim 70 \%$ of $\phi_{\nu_e}$ and $\sim 30 \%$ of $\phi_{\bar{\nu}_x}$.
The average energy of $\nu_e$ is about 10 MeV, so that the large
ratio is obtained.
As the second pattern [pattern (B)], the ratio is $r_{{\rm L/H}} \sim 0.4$ 
in the inverted mass hierarchy and nonadiabatic MSW-H resonance.
The main component of the $\bar{\nu}_e$ flux is originating from
$70 \%$ of $\bar{\nu}_e$.
The average energy of $\bar{\nu}_e$ is about 15 MeV, so that 
the lower component is smaller than the former case.
The third pattern [pattern (C)] corresponding to the other two cases 
indicates the ratio of about 0.25.
The neutrino signal is almost identical to the original signal of
$\nu_{\mu,\tau}$ or $\bar{\nu}_{\mu,\tau}$.
The low energy component of the spectrum is smaller than those of
$\nu_e$ and $\bar{\nu}_e$.

We see a dip of the ratios at $t \sim 1 - 2$ s in the case of
the inverted mass hierarchy and the adiabatic MSW-H resonance.
This is due to the fact that the shock wave passes across the density
region of RSF-H resonance at that time.
The density gap produced by the shock propagation changes the adiabaticity
of RSF-H resonance.

We consider the case of $Y_e = 0.51$ (see the right-hand panel of Fig. 9).
We see two patterns of the time evolution of the ratios, 
$r_{{\rm L/H}} \sim$ 0.4 and 0.25.
The large value of the ratio [pattern (B)] is shown in the normal mass 
hierarchy.
The neutrino signal has been converted from $\bar{\nu}_e$ by about 70 \% and
from $\bar{\nu}_{\mu,\tau}$ by about 30 \%.
This is the same as in the case of the inverted mass hierarchy,
nonadiabatic MSW-H resonance, and $Y_e = 0.49$.
The small ratio [pattern (C)] corresponds to the inverted mass hierarchy.
Therefore, we see a clear dependence on mass hierarchy in the event ratio
but do not see the dependence on $\sin^22\theta_{13}$.
We also see a small dip as the shock propagation effect of RSF-H resonance
at $t \sim 0.5$ s.

Thus, we can classify the event ratios $r_{{\rm L/H}}$ into 
three patterns:
(A) large ratios, (B) intermediate ratios, and (C) small ratios.
The classified patterns of $r_{{\rm L/H}}$ are listed in Table I.

\section{Discussion}

\subsection{Time evolution of neutrino signal}

We have shown that the SN neutrino signal strongly depends on
electron fraction in the innermost region of the SN ejecta
taking account of the RSF conversion.
Here we discuss how the neutrino signal changes with time when we consider
time evolution of the electron fraction.
Recently, Arcones et al. \cite{aj07} calculated a long time ($\sim 10$ s)
evolution of neutrino-driven outflows in SNe.
In their result, the electron fraction exceeds 0.5 just after the core
bounce and reaches the maximum value of 0.52.
Then, it gradually falls off and goes below 0.5 at $\sim 2 - 3$ s.
After this time it stays about 0.48 up to $\sim 10$ s.
The electron fraction is almost homogeneous
in the outflow region.
They showed small dependence on the progenitor masses.

We consider the time evolution of the electron fraction in the innermost
region by mimicking their result \cite{aj07}.
We set the outer edge of the innermost region to be $M_r = 1.43 M_\odot$
as explained in Sec. II, and the electron fraction is set to be constant 
in the innermost region.
At $t =$ 0, 0.5, 1, and 2 s, we set the electron fraction to be 0.49, 0.51,
0.52, and 0.51, respectively.
At $t = 3$ s, we adopt two $Y_e$ values of 0.505 and 0.495.
In the case of $Y_e = 0.5$, the RSF-H and RSF-L resonances disappear
as explained in Sec. III below Eqs. (18) and (19).
After 4 s, we set the electron fraction as 
\begin{equation}
Y_e = 0.49 - 0.005 (t-4) .
\end{equation}

First, the time evolution of the $\bar{\nu}_e$ event number ratio 
$r_{{\rm L/H}}$ with the water-{\v C}erenkov detector is shown in Fig. 10.
Error bars are evaluated assuming that the event number rate
$dN_i/dt$ has an uncertainty of $\sqrt{dN_i/dt}$.
In the normal mass hierarchy, we see an enhancement of the event ratio
$r_{{\rm L/H}} \sim 0.4$ in $t \sim 1-3$ s compared to the event ratio
$r_{{\rm L/H}} \sim 0.25$ in the other time.
During $t \sim 1 - 3$ s the electron fraction is larger than 0.5 and the flavor
conversion at RSF-H resonance is different from the case in $Y_e < 0.5$.
We do not see a clear dependence of $r_{{\rm L/H}}$ on the adiabaticity of
MSW-H resonance.

\begin{figure*}
\includegraphics[width=6.7cm]{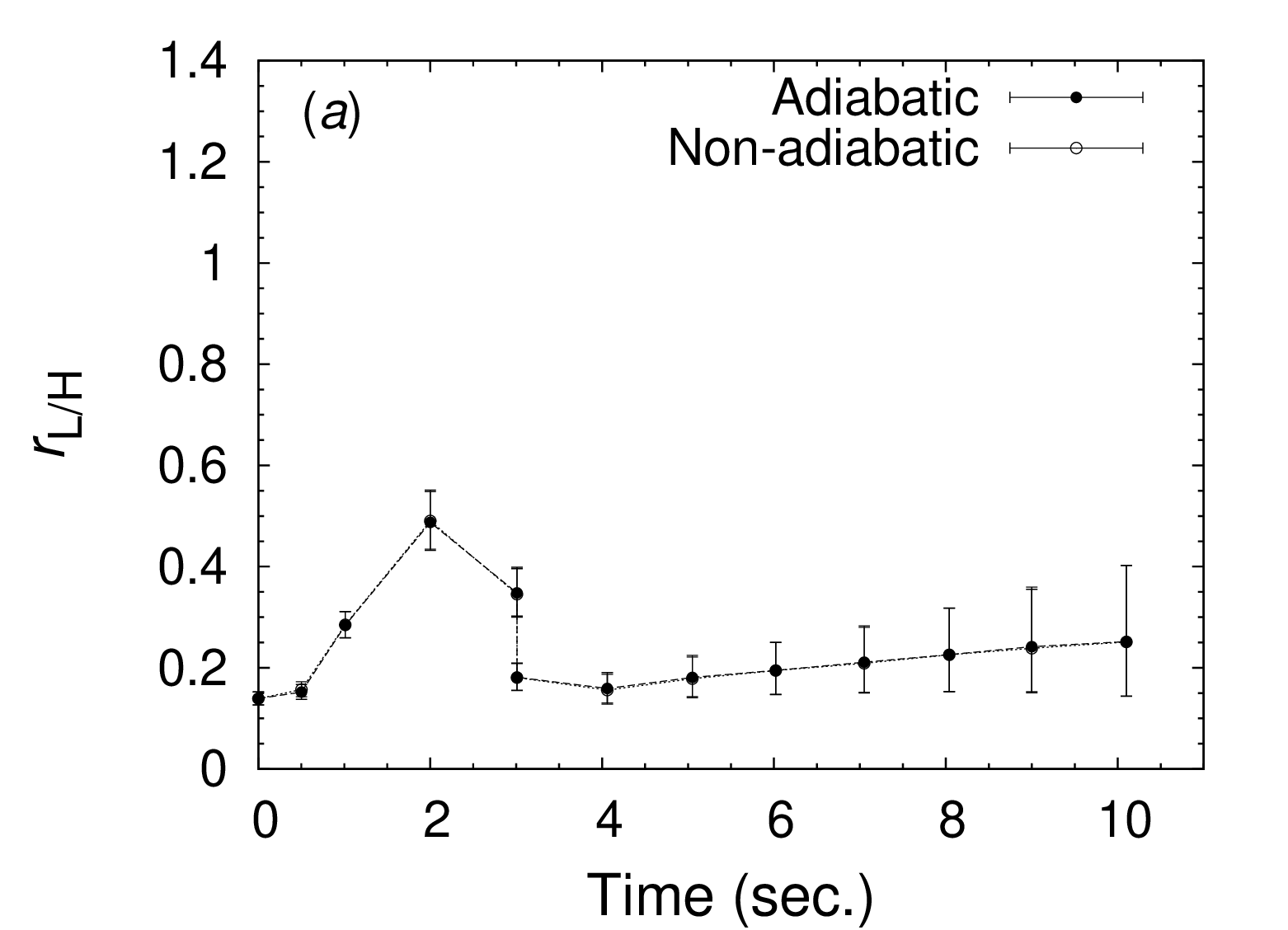}
\includegraphics[width=6.7cm]{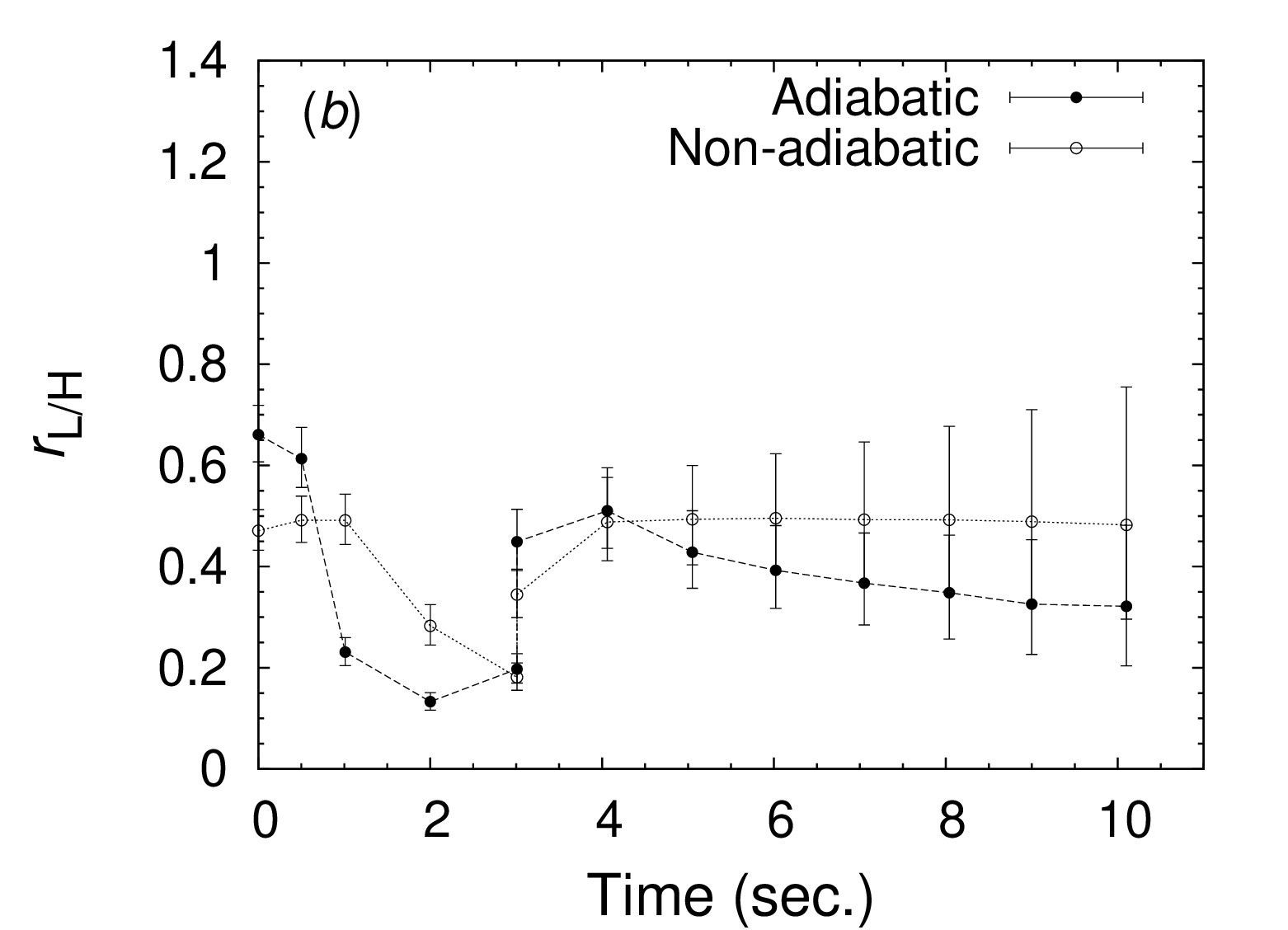}
\caption{\label{fig:rLH_tevol}
Same as Fig. 10 but for the initial neutrino spectra of the Livermore
model: Fermi distributions with
($T_{\nu_e}$, $T_{\bar{\nu}_e}$, $T_{\nu_{\mu,\tau}}$) =
(3.5, 4, 7) MeV and
($\mu_{\nu_e}$, $\mu_{\bar{\nu}_e}$, $\mu_{\nu_{\mu,\tau}}$) =
(7.4, 10, 0) MeV.
}
\end{figure*}

In the inverted mass hierarchy, a dip in the event ratio 
$r_{{\rm L/H}} \sim 0.25$ appears in $t \sim 1-3$ s.
This dip reflects the adiabatic RSF-H conversion in the case of $Y_e > 0.5$.
The event ratio $r_{{\rm L/H}}$ does not change with time very much
after $t = 4$ s.
It depends on the adiabaticity of the MSW-H resonance.
In the adiabatic case the event ratio is about $\sim 0.5 - 0.6$.
In the nonadiabatic case it is about $0.4$.
Even considering an uncertainty of the event rate, the event ratio
would indicate the adiabaticity of the MSW-H resonance.

We note that even if the RSF conversion does not contribute to flavor changes,
the enhancement of $r_{{\rm L/H}}$ will be seen.
If mass hierarchy is inverted and the MSW-H resonance is partly adiabatic
($\sin^22\theta_{13} \sim 10^{-3}$), the event ratio $r_{{\rm L/H}}$
increases by the shock passage of the MSW-H resonance 
(e.g., \cite{tk04,fl05,kk08}).
Therefore, it is difficult to specify the cause of the enhancement of
the $r_{{\rm L/H}}$ ratio.

We discuss observable effects of the RSF conversion taking into account 
uncertainties of the SK detector resolution and efficiency, the initial 
neutrino spectra, and the dependence on the distance from a SN.
We have considered the uncertainties by the detector resolution and
efficiency of SK-III for the event rate ratio $r_{{\rm L/H}}$.
First, we discuss the influence of the difference in the detector resolution 
and efficiency of the second phase of the SK experiment (SK-II).
The energy resolution function of SK-II is represented by
$R(E_e,E_e') = 0.0536+0.52 \sqrt{E_e}+0.458 E_e$.
The detection efficiency is considered to be unity for $E_e \ge 7$ MeV
and otherwise zero.
We do not find a drastic influence of the different energy resolution and 
efficiency.
The value of $r_{{\rm L/H}}$ systematically decreases by 6\%$-$9 \% 
throughout the time evolution in the normal mass hierarchy and in the 
inverted mass hierarchy with $\sin^22\theta_{13} = 1 \times 10^{-6}$.
In the case of the inverted mass hierarchy with $\sin^22\theta_{13} = 0.04$,
the $r_{{\rm L/H}}$ value decreases by about 15 \%, except for 
$t \sim 1 - 3$ s where the decrease in the $r_{{\rm L/H}}$ is similar to
the case in the normal mass hierarchy.
These differences mainly arise from the shift of the energy threshold
in the detection efficiency from 5 to 7 MeV.

\begin{figure*}
\includegraphics[width=6.7cm]{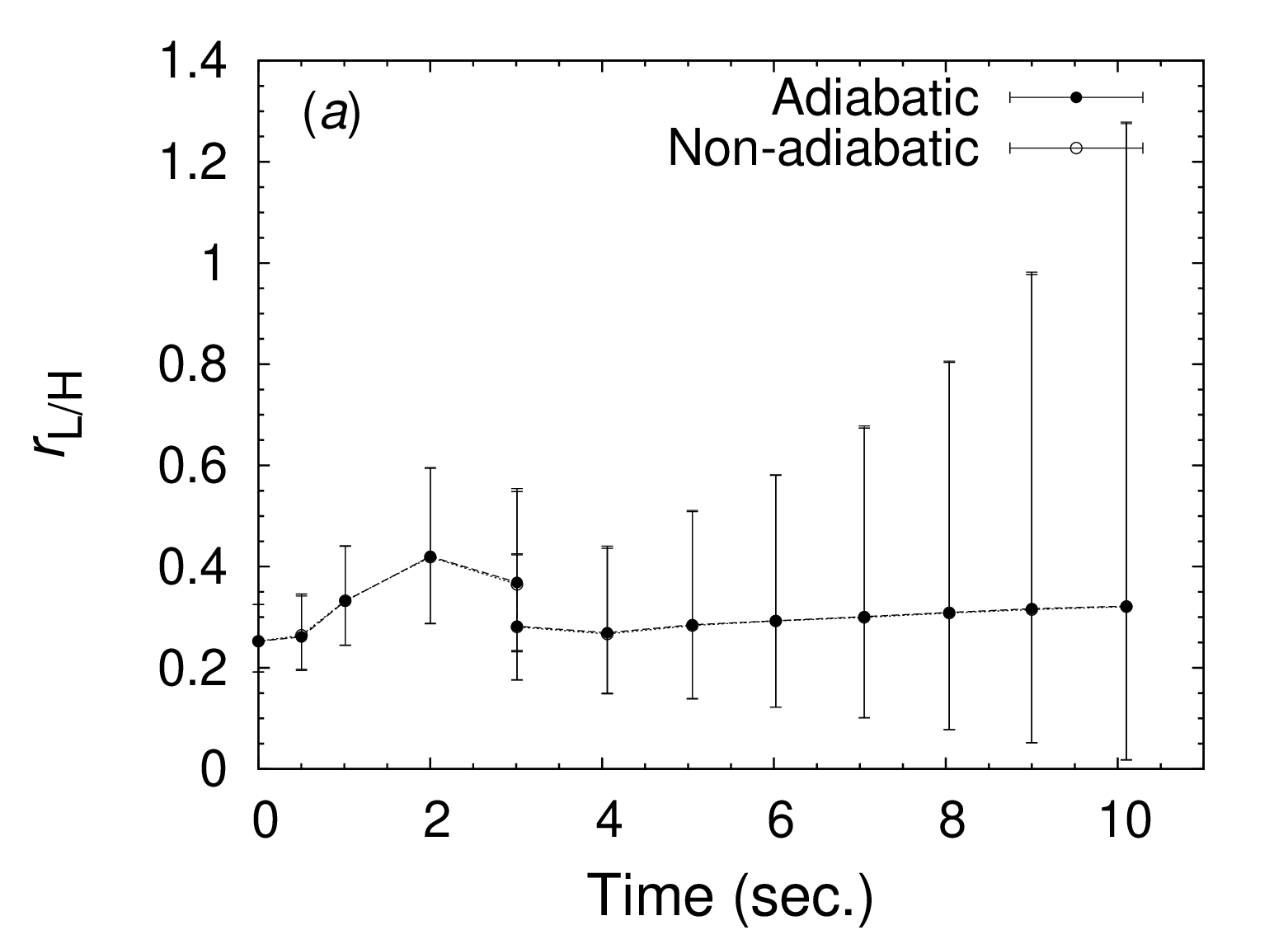}
\includegraphics[width=6.7cm]{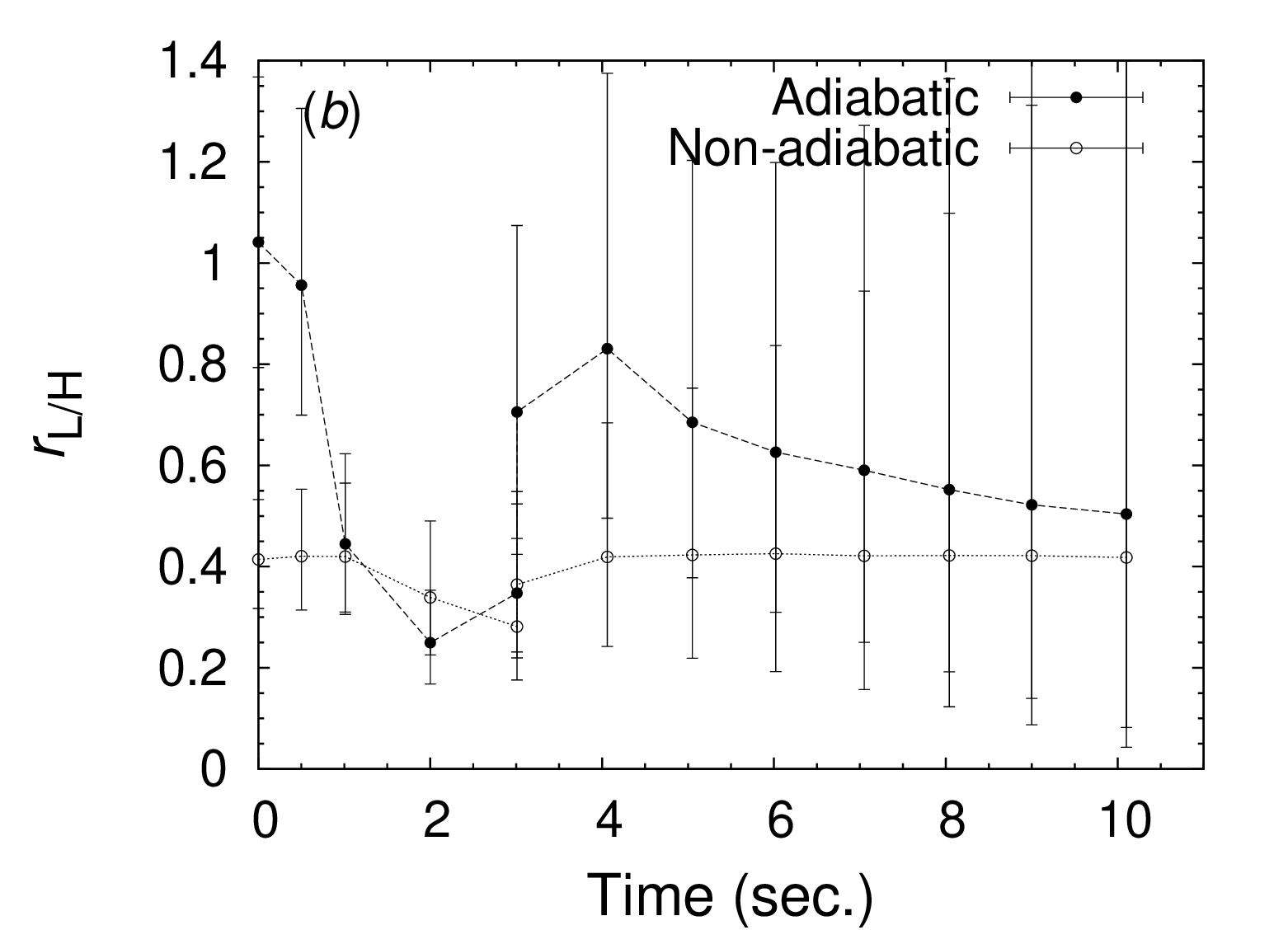}
\caption{\label{fig:rLH_tevol}
Same as Fig. 10 but for the distance from a SN of $d = 30$ kpc.
}
\end{figure*}

We secondly discuss the dependence on the initial neutrino spectra using 
those corresponding to a SN simulation of the Livermore group \cite{ts98}.
Although we used the neutrino temperatures of
($T_{\nu_e}$, $T_{\bar{\nu}_e}$, $T_{\nu_{\mu,\tau}}$) = 
(3.2, 5, 6) MeV and zero chemical potentials, we here adopt 
the temperatures of
($T_{\nu_e}$, $T_{\bar{\nu}_e}$, $T_{\nu_{\mu,\tau}}$) = 
(3.5, 4, 7) MeV and the chemical potentials of
($\mu_{\nu_e}$, $\mu_{\bar{\nu}_e}$, $\mu_{\nu_{\mu,\tau}}$) = 
(7.4, 10, 0) MeV in the Livermore model.
This spectrum set indicates smaller temperature difference between
$\nu_e$ and $\bar{\nu}_e$ and larger temperature difference
between $\nu_{\mu,\tau}$ and $\bar{\nu}_e$.

Figure 11 shows the time evolution of the event rate ratio in the Livermore
model.
In the normal mass hierarchy [Fig. 11(a)], the enhanced $r_{{\rm L/H}}$ 
ratio is 0.49 at $t = 2$ s, whereas the ratio is $0.14 \sim 0.25$ in 
$t = 0 \sim 0.5$ s and $3 \sim 10$ s.
The enhancement in this model is seen clearer than in our model.
In the inverted mass hierarchy with $\sin^22\theta_{13} = 0.04$ 
[adiabatic case in Fig. 11(b)], the $r_{{\rm L/H}}$ ratio is about 0.6 
and decreases to 0.13 at $t = 2$ s.
In the case of $\sin^22\theta_{13} = 1 \times 10^{-6}$ 
[nonadiabatic case in Fig. 11(b)], the $r_{{\rm L/H}}$
ratio decreases from 0.49 ($t = 0.5$ and 1 s) to 0.18 
($t = 3$ s, $Y_e > 0.5$).
The decrease in $r_{{\rm L/H}}$ is seen even in the case of such a small 
$\sin^22\theta_{13}$ value.
This leads to smaller $\theta_{13}$ dependence in the neutrino signal.
These changes seen in Fig. 11 from Fig. 10 are explained by the fact that
the temperature difference between $\nu_{\mu,\tau}$ and
$\bar{\nu}_e$ is larger in the Livermore model.
Thus, the evidence for the RSF conversion could be observed even taking 
account of the uncertainties on detector and the initial neutrino spectra.

Thirdly, let us discuss the dependence on the distance from a SN.
If a SN explodes at a longer distance than 10 kpc, the signal of the RSF 
conversion in the event rate ratio may be weaker for a smaller event rate.
We show the event ratio in the case of a SN at the distance of 30 kpc
in Fig. 12.
In the normal mass hierarchy [Fig. 12(a)], it is difficult to identify
the enhancement of $r_{{\rm L/H}}$ observationally from $t = 0.5$ s to 3 s
because the error bars overlap.
In the case of the inverted mass hierarchy with $\sin^22\theta_{13} = 0.04$
[Fig. 12(b)], on the other hand, the reduction of $r_{{\rm L/H}}$ from
$t = 0$ s to 2 s is still detectable.
Therefore, we could observe the decrease in the event rate ratio of the
neutrinos from a SN in our Galaxy.
A megaton size detector such as Hyper-Kamiokande could observe the
evidence for the RSF conversion in neutrino signal from a SN at the
distance of $\sim 100$ kpc which well includes Large Magellanic Cloud
and Small Magellanic Cloud.

We evaluated the time evolution of $\bar{\nu}_e$ events by 
$p(\bar{\nu}_e,e^+)n$.
Neutrino events also occur through other neutrino interactions with
$e^-$ and $^{16}$O.
The time variation of $\nu_e$ events is also useful to investigate 
the RSF conversion.
If the RSF conversion is effective in a SN, time variations as their traces
should be observed in both $\nu_e$ and $\bar{\nu}_e$ spectra.
If the RSF conversion is ineffective and the shock propagation changes
the MSW effect, however, time variation will be observed in only one of
the spectra.
Therefore, the observation of time evolution of both of the $\nu_e$ and
$\bar{\nu}_e$ spectra will be a good test for the occurrence of 
the RSF conversion.
We can consider $\nu_e$ events through $^{16}$O($\nu_e,e^-)^{16}$F reaction
\cite{ha87}.

In the future, a gadolinium trichloride (GdCl$_3$) water-{\v C}erenkov 
detector is expected to successfully pick out the $\bar{\nu}_e$ events by 
$p(\bar{\nu}_e,e^+)n$ and $^{16}$O($\bar{\nu}_e,e^+n)^{15}$N from 
the neutrino events owing to neutron detection by Gd \cite{bv04}.
Events by electron-neutrino scattering will also be distinguished by 
forward-peaked angular distribution.
After the reaction distinction, time variations of $\nu_e$ events by the
$^{16}$O reaction as well as $\bar{\nu}_e$ events by $p(\bar{\nu}_e,e^+)n$
will be successfully evaluated.
New findings by megaton-size water-{\v C}erenkov detectors have been 
discussed in \cite{ky08}.
When such large neutrino detectors are established,
$\sim 10^6$ events of $p(\bar{\nu}_e,e^+)n$ and $\sim 10^4$ events of 
$^{16}$O are expected from the SN.
Large events of $^{16}$O reactions will clarify $\nu_e$ spectrum
as well as $\bar{\nu}_e$ spectrum.

\subsection{Magnetic field strength}

In this study, we set the magnetic field of $B_0 = 1 \times 10^{11}$ G.
We assumed that the magnetic field decreases proportional to $r^{-3}$, 
so that the magnetic field at the RSF-H resonance is smaller than $B_0$.
The location of the RSF-H resonance and the corresponding magnetic field
are $r = 5 \times 10^8$ cm and $B = 8 \times 10^8$ G, respectively, 
at the presupernova stage.
They are $r = 8 \times 10^8$ cm and $B = 2 \times 10^8$ G
at $t = 4$ s after the explosion.
Here we discuss the magnetic field at the RSF-H resonance.

We roughly estimate the adiabaticity of the RSF-H resonance.
The adiabaticity parameter of the RSF-H resonance $\gamma_{{\rm RSF-H}}$ is
described as (e.g. \cite{af03})
\begin{equation}
\gamma_{{\rm RSF-H}} \sim 
\frac{8 E_\nu}{\Delta m^2_{31}}
(\mu_{e\tau}B_{\bot})^2
\left| \frac{d \ln (\rho(1-2Y_e))}{dr} \right|^{-1}_{{\rm RSF-H}} \quad .
\end{equation}
The adiabaticity is proportional to $B_{\bot}^2$.
We obtained that the adiabaticity of the RSF-H resonance in this study
is about 10.
We evaluate the survival probability of $\bar{\nu}_e$ in the SN
ejecta as a function of the transverse magnetic field strength.
Here we assumed the normal mass hierarchy, $\sin^22\theta_{13} = 0.04$, and
the neutrino energy of 20 MeV.
We used the SN density profile at $t = 4$ s and set the electron fraction
in the innermost region of 0.49.
Figure 13 shows the survival probability of $\bar{\nu}_e$ with the relation
to the magnetic field.
The survival probability of $\bar{\nu}_e$ should be 0.01 and 0.71 when
the RSF-H resonance is adiabatic and nonadiabatic, respectively.
We see that the resonance is adiabatic in the case of $B_0 > 5 \times 10^{10}$
G.
The adiabatic condition for the magnetic field at the RSF-H resonance is
$B > 4 \times 10^8$ G at the presupernova stage and 
$B > 1 \times 10^8$ G at $t = 4$ s after the explosion.

\begin{figure}[b]
\includegraphics[angle=-90,width=6.7cm]{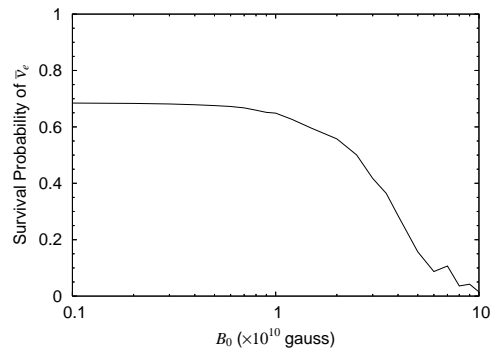}
\caption{\label{fig:prob_bdep}
The survival probability of $\bar{\nu}_e$ as a function of the transverse
magnetic field strength $B_0$.
Detailed information of the SN density profile and neutrino
oscillation parameters is explained in the text.
}
\end{figure}

We note that the adiabatic condition of the neutrino magnetic moment
$\mu_{e\tau}$ is derived from the proportionality of $\gamma_{{\rm RSF-H}}$
to ($\mu_{e\tau}B)^2$.
The RSF-H resonance in this study is adiabatic in the case of
$(\mu_{e\tau}/\mu_B)B_0 > 5 \times 10^{-2}$ G.
Since we set $B_0$ to be $1 \times 10^{11}$ G, the adiabatic condition of 
the RSF-H resonance on the neutrino magnetic moment is 
$\mu_{e\tau} > 5 \times 10^{-13} \mu_B$.
Thus, an effect of the RSF conversion could be observable in the
neutrino spectra if the neutrino magnetic moment is larger than
$5 \times 10^{-13} \mu_B$.
We should also note that this constraint strongly depends on the model
of the magnetic field in the RSF-H resonance.

The evolution of the magnetic field distribution of massive stars was
estimated in \cite{hw05}.
Azimuthal magnetic moment $B_\phi$ is dominated in radiative layer.
The magnetic field $B_\phi$ is roughly constant in carbon core at
carbon ignition.
The magnetic field becomes larger than $B_\phi \sim 10^7$ G.
It becomes $5 \times 10^7$ G in the Si depletion stage and 
$5 \times 10^9$ G at the presupernova stage.
The magnetic field at the RSF-H resonance in our model is between the two
values of the magnetic field.
Therefore, the RSF-H resonance would be adiabatic if the transition magnetic
moment is on the order of $10^{-12} \mu_B$.

We note that small dependence of the azimuthal magnetic field in carbon
core may make the RSF-X and RSF-L resonances adiabatic.
The resonance densities of the RSF-X and RSF-L resonances are larger than
the resonance density of the MSW-H resonance, and these resonances are inside
the carbon core.
If the magnetic field of the carbon core is $\sim 10^7$ G or more,
flavor change by the RSF-X and/or RSF-L resonances may occur.
On the other hand, if the magnetic field is different among individual
layers, the RSF-X and RSF-L resonances would be nonadiabatic because of
small magnetic field.

We also note that the adiabaticities of the RSF resonances would depend on 
the neutrino emission angle from the rotation axis.
If azimuthal magnetic field is dominated, the RSF conversion is insensitive
to the neutrino emission direction.
On the other hand, if dipole magnetic field is dominated and the axes
of rotation and dipole magnetic field are the same, the RSF conversion 
strongly depends on the neutrino emission direction.
The neutrinos emitting to the equatorial direction are affected by
the RSF conversion most effectively.

Neutron stars having very strong magnetic field $B \sim 10^{14}$ G
are called magnetars.
Although magnetar formation is still unsolved, magneto-driven SN is considered
to be a promising formation process.
Magnetohydrodynamical simulations of such SNe have been performed
(e.g., \cite{ks04,ky05,tk09}).
When a rotating massive star collapses, a very strong magnetic
field is formed along the rotational axis and a jet is launched.
The magnetic field becomes larger than $\sim 10^{14}$ G in the jet.
If the density in the polar region becomes low after the jet and
the magnetic field is still strong, the RSF conversion could change
neutrino flavors.
However, the flavor change may be different from our proposition because
the structure of the density and electron fraction can be quite different
from our model.

\subsection{Effect of neutrino-neutrino interaction}

Neutrino flavor change by neutrino-neutrino interaction is now one of
the hot topics in the study of neutrino oscillation.
Neutrino flux in hot-bubble and wind regions of a SN is so large that
the effect of the neutrino-neutrino interaction potential becomes larger
than that of the neutrino-electron interaction.
The final neutrino energy spectra changed by this interaction strongly 
depend on the initial neutrino energy spectra and neutrino flux 
(e.g., \cite{df06,fl07,gv08,fl09,dd09}).
The flavor change occurs in the radius of $\sim 10^7$ cm, i.e., in more inner 
region than the RSF-H resonance.
Thus, the neutrino energy spectra after the neutrino-neutrino interaction
can be treated as the initial neutrino spectra for the RSF conversion.

The flavor change by neutrino-neutrino interaction is complicated, so that
it would be quite difficult to expect the final neutrino energy spectra taking
into account the neutrino-neutrino interaction and the RSF conversion.
However, we can expect the final neutrino energy spectra in the case of
classical swap (e.g., \cite{fl07,gv08,fl09}).
In normal mass hierarchy there is no effect by neutrino-neutrino interactions.
In this case our result is used without any modifications.
In inverted mass hierarchy, neutrino energy spectra split at a definite energy
and antineutrino energy spectra swap or split at a smaller energy.
If the RSF-H resonance is adiabatic, the $\bar{\nu}_e$ spectrum shifts to 
the low energy side in $Y_e > 0.5$, and the enhancement of $r_{{\rm L/H}}$ 
will be seen in the time variation of the neutrino event.
On the other hand, the shock effect on the MSW effect will be seen
as the reduction of $r_{{\rm L/H}}$ when the classical swap
is taken.
Therefore, the effect of the RSF conversion will be distinguished from
the MSW effect.

\section{Summary}

The electron fraction $Y_e$ becomes larger than 0.5 in several seconds 
in the innermost region including the location of the RSF-H resonance 
in SN ejecta.
We investigated the RSF conversion effects of SN neutrinos on the electron
fraction in the SN ejecta.
The obtained results and discussions are summarized as follows.
\begin{enumerate}
\item
The converting flavors in the RSF-H resonance are different between 
$Y_e < 0.5$ and $Y_e > 0.5$ in the RSF-H resonance region.
In a normal (an inverted) mass hierarchy case, the flavor conversion
occurs at the RSF-H resonance for $\bar{\nu}_e \leftrightarrow \nu_{\mu,\tau}$
($\nu_e \leftrightarrow \bar{\nu}_{\mu,\tau}$) in the case of $Y_e < 0.5$
and for $\nu_e \leftrightarrow \bar{\nu}_{\mu,\tau}$
($\bar{\nu}_e \leftrightarrow \nu_{\mu,\tau}$) in the case of $Y_e > 0.5$.

\item
When there is a region of $Y_e > 0.5$ including the RSF-H resonance, the ratio
of low energy component to high energy component of the neutrino event
shows a trend opposite to the one in the case of the RSF-H resonance
with $Y_e > 0.5$.

\item
Detailed simulations of SN explosions have indicated that the
electron fraction in the innermost region becomes larger than 0.5 in 
a few seconds and it becomes smaller than 0.5 afterwards.
In the normal mass hierarchy, the energy ratio $r_{{\rm L/H}}$
slightly enhances in the period of $Y_e > 0.5$ and 
it changes to a small constant value afterwards.
In an inverted mass hierarchy, $r_{{\rm L/H}}$ becomes small first
and it changes to a large constant value.

\item
The adiabaticity of the RSF-H resonance is proportional to
$(\mu_{e\tau}B_\bot)^2$.
Azimuthal magnetic field at the presupernova stage would be large
enough to make the RSF-H resonance adiabatic.
The magnetic moment of $\mu_{e\tau} > 5 \times 10^{-13} \mu_B$ would
produce an observable effect on the RSF conversion when we set
$B_0$ equal to $1 \times 10^{11}$ G.

\end{enumerate}

\begin{acknowledgments}
TY would like to thank Koichi Iwamoto for helpful discussions on
the numerical scheme for neutrino oscillations.
We are indebted to Naotoshi Okamura for fruitful comments.
Numerical computations were in part carried out on the general-purpose
PC farm at Center for Computational Astrophysics, CfCA, of
National Astronomical Observatory of Japan.
This work has been supported in part by the Grant-in-Aid for Scientific 
Research [(C)20540284, (A)20244035] and Scientific Research on Innovative 
Areas (20105004) of the Japanese Ministry of Education, Culture, Sports, 
Science, and Technology.
\end{acknowledgments}





\end{document}